\documentclass[12pt,a4paper]{article}
\usepackage{amssymb,amsfonts,amsmath,amsthm,euscript}
\usepackage[latin1]{inputenc}
\usepackage[english]{babel}
\usepackage[left=2.7cm,right=2.7cm,bottom=2.67cm,top=2.67cm]{geometry}
\usepackage{dsfont}
\usepackage{mathrsfs}
\usepackage{enumerate}
\usepackage{multicol}
\usepackage{multirow}
\usepackage{graphicx}
\usepackage{subfigure}
\usepackage{color}
\usepackage{url}
\usepackage{natbib}
\usepackage{placeins}
\usepackage{slashbox}

\newcommand{\R}{\mathds{R}}
\newcommand{\C}{\mathds{C}}
\newcommand{\N}{\mathds{N}}
\newcommand{\Z}{\mathds{Z}}
\newcommand{\bs}[1]{\boldsymbol{#1}}
\newcommand{\F}{\mathscr{F}}
\newcommand{\E}{\mathds E}
\newcommand{\ga}{\Gamma}
\newcommand{\var}{\mathrm{Var}}
\newcommand{\df}[2]{\frac{\partial #1}{\partial #2}}
\newcommand{\tr}{\mathrm{tr}}
\theoremstyle{plain}
\newtheorem{theorem}{Theorem}[section]
\theoremstyle{definition}
\newtheorem{remark}{Remark}[section]

\long\def\sfootnote[#1]#2{\begingroup%
\def\thefootnote{\fnsymbol{footnote}}\footnote[#1]{#2}\endgroup}

\def\bfootnote{\xdef\@thefnmark{}\@footnotetext}

\begin{document}
\thispagestyle{empty}
{\centering
\Large{\bf Beta autoregressive fractionally integrated moving average models}\vspace{.5cm}\\
\normalsize{ {\bf Guilherme Pumi$\!\phantom{i}^{\mathrm{a,}}$\sfootnote[1]{Corresponding author.}\let\thefootnote\relax\footnote{\hskip-.3cm$\phantom{s}^\mathrm{a}$Mathematics and Statistics Institute - Universidade Federal do Rio Grande do Sul - 9500,  Bento Gon\c calves Avenue - 91509-900, Porto Alegre - RS - Brazil.
}, Marcio Valk$\!\phantom{i}^{\mathrm{a}}$, Cleber Bisognin$\!\phantom{i}^{\mathrm{b}}$, F\'abio Mariano Bayer$\!\!\phantom{s}^\mathrm{b}$\let\thefootnote\relax\footnote{\hskip-.3cm$\phantom{s}^\mathrm{b}$ Departamento de Estat\'istica and LACESM - Universidade Federal de Santa Maria, Santa Maria - RS - Brazil.
}  and Taiane Schaedler Prass$\!\!\phantom{s}^\mathrm{a}$  \\
\let\thefootnote\relax\footnote{E-mails: guilherme.pumi@ufrgs.br, marciovalk@gmail.com, cbisognin@ufrgs.br, bayer@ufsm.br and taianeprass@gmail.com}}\\
\vskip.3cm
}}

\begin{abstract}
In this work we introduce the class of beta autoregressive fractionally integrated moving average models for continuous random variables taking values in the continuous unit interval $(0,1)$.
The proposed model accommodates a set of regressors and a long-range dependent time series structure. We derive the partial likelihood estimator for the parameters of the proposed model, obtain the associated score vector and Fisher information matrix.
We also prove the consistency and asymptotic normality of the estimator under mild conditions.
Hypotheses testing, diagnostic tools and forecasting
are also proposed.
A Monte Carlo simulation is considered to evaluate the finite sample performance of the partial likelihood estimators and to study some of the proposed tests.
An empirical application is also presented and discussed.\vspace{.2cm}
\noindent \textbf{MSC2000 subject classification:} 62M10 $\cdot$ 62F12 $\cdot$ 62J12 $\cdot$ 62J99.
 %62J99  	Linear inference, regression - None of the above, but in this section
 %62M10  	Time series, auto-correlation, regression, etc.
 %62F12     Asymptotic properties of estimators
 %62J12  	Generalized linear models
\end{abstract}

\section{Introduction}

In this work we are interested in time series whose values are restrained
to a continuous interval of the real line, say $(a,b)$, where $a<b$.
Without loss of generality we consider series in the unit interval $(0,1)$.
One typical broad case is when the time series represent rates and proportions observed over time. % among many others.
Building over the works of \cite{Zeger1988}, \cite{Benjamin2003} and \cite{Ferrari2004},  \cite{Rocha2009} introduces
the class of beta autoregressive moving average models ($\beta$ARMA),
which can be viewed as a specialization of the
generalized autoregressive moving average models (GARMA) \citep{Benjamin2003}
for beta distributed variates.
%
%The approach of GARMA models is based on including time dependent structure into a generalized linear models (GLM) framework \citep{McCullagh1989}
%for which a term following an autoregressive moving average model
%%of order $p$ and $q$ (ARMA$(p,q)$)
%(ARMA) \citep{Brockwell1991} is added in the regression structure. In the same sense,
%$\beta$ARMA model introduces an ARMA structure in the beta regression model \citep{Ferrari2004},
%being a dynamic model for continuous random variables assuming values in the standard unit interval $(0, 1)$.
Applications of the  $\beta$ARMA model spam over several areas, such as medicine \citep{Zou2010}, online monitoring \citep{Guolo2014}, neuroscience \citep{Wang2012}, among many others.
%
%\cite{Cox1981} divided the analysis of non-Gaussian time series into two main categories: the observation-driven and the parameter-driven approaches.
%The so-called parameter-driven approaches comprehend models for which a key component is the presence of a latent variable and include bayesian and state space modeling. An account of developments in the 90's can be found in \cite{Durbin2000}. For a parameter-driven approach in the context of beta regression, see \cite{daSilva2011}. This work is concerned with the observation-driven approach, as are \cite{Zeger1988, Benjamin2003,Ferrari2004,Rocha2009}.
%An account regarding the observation-driven  approach can be found in \cite{Fahrmeir2001}.

Let $\{y_t\}_{t=1}^\infty$ be a process of interest and, aiming towards prediction and the use of partial likelihood inference, let $\bs{x}_{t-1}'$ denote the $l$-dimensional vector of  (exogenous random) covariates at time $t-1$ and any non-random component up to time $t$, to be considered in the model (a possible intercept will be considered in the model separately). Let $\F_{t-1}$ denote the $\sigma$-field generated by the past and present (when known) explanatory variables and possibly past values of the response variable, if they are included in the model. In this framework, the $\sigma$-field $\F_{t-1}$ represents all the observer's knowledge about the model up to time $t-1$, with a possible addition of predetermined variables at time $t$.
%This exceptionality of knowing the occurrence of a variable at time $t$, in a previous state $t-1$, occurs, for instance, if the model includes some deterministic feature such as dummy variables, an harmonic component (to deal with a seasonality), a deterministic trend, etc.

{Inference in  the context of $\beta$ARMA process \citep{Rocha2009} is conducted using a conditional likelihood approach, which only allows for deterministic covariates to be introduced in the model. In this work, we adopt the more general  approach of partial likelihood, which allows for $\bs{x}_{t-1}'$ to contain deterministic covariates, as in the conditional likelihood approach, but also enables the inclusion of (time dependent) random covariates, as well as any type of interaction or a mixture of these.} For further details on partial likelihood inference we refer the reader to \cite{Cox1975}, \cite{Wong1986} and \cite{Jacod1987,Jacod1990}. For details on partial likelihood in time series following generalized linear models, we refer to \cite{Fokianos2004,Kedem2002} and references therein.

%As for the random component, it comprehends the specification of the response's conditional distribution given the observed past. We shall
This work is concerned with an observation-driven model in which the random component follow a conditional beta distribution, parameterized as \citep{Ferrari2004}:
\begin{align}\label{p1}
f(y_t;\mu_t,\nu|\F_{t-1})=\frac{\ga(\nu)}{\ga(\nu\mu_t)\ga\big(\nu(1-\mu_t)\big)}\,y_t^{\nu\mu_t-1}(1-y_t)^{\nu(1-\mu_t)-1},
\end{align}
for $0<y_t<1$,
$0<\mu_t<1$,
and
$\nu>0$,
where $\E(y_t|\F_{t-1})=\mu_t$ and $\var(y_t|\F_{t-1})=\frac{\mu_t(1-\mu_t)}{1+\nu}$.
We note that $\nu$ is a precision parameter in the sense that the greater the $\nu$, the smaller the variance of the distribution.
The systematic component follows the usual approach of GLM with an additional dynamic term.
Let $g(\cdot)$ be a twice differentiable monotonic one-to-one link function for which the inverse link is of class $\mathcal{C}^2(\R)$ (the class of twice continuously differentiable functions in $\R$).
Consider the additive specification
\begin{align}\label{e:gen}
g(\mu_t)={\eta}_t=\bs{x}_{t-1}'\bs{\beta}+\tau_t,
\end{align}
where $\bs{\beta}'=(\beta_1,\dots,\beta_l)$ are the coefficient related to the covariates
and ${\eta}_t$ is the linear predictor.
The particular form of $\tau_t$ is discussed in several papers \citep{Benjamin2003,Rocha2009,Fokianos2004}.
In $\beta$ARMA models \citep{Rocha2009},
$\tau_t$ is assumed to follow an ARMA$(p,q)$ process of the type
\begin{align*}%\label{p2}
\tau_t=\sum_{j=1}^p\phi_j\big(g(y_{t-j})-\bs{x}_{t-j-1}'\bs{\beta}\big)+\sum_{k=1}^q\theta_kr_{t-k},
\end{align*}
where $p$ and $q$ and $\bs{\phi}=(\phi_1,\dots,\phi_p)'$ and $\bs{\theta}=(\theta_1,\dots,\theta_q)'$ denote the order and coefficients of the autoregressive and moving average parts of the model, respectively, and $r_{t}$ denotes an error term. When $\bs{x}_{t-1}$ is non-random, this is the $\beta$ARMA model of \cite{Rocha2009}.%\footnote{recall that, in our framework, $\bs{x}_{t-1}$ contains all non-random components up to time $t$}.

An advantage of specification \eqref{e:gen} is that even though the conditional mean is transformed, it is actually $\mu_t$ that is being modeled. In some applications it is common to model $g(y_t)$ and then transform estimates  back by applying $g^{-1}$ which can be problematic (Jensen's inequality, delta method, etc.).
Observe that the time series part of the $\beta$ARMA model can only accommodate short range dependence, structure that may not be enough in certain situations. In this work we propose a generalization of the $\beta$ARMA model of \cite{Rocha2009}  by allowing $\tau_t$ to accommodate long-range dependence.
%
%
% more general types of dependence, such as long-range dependence \citep{Brockwell1991,palma2007}, which is characterized by a very slow covariance decay as the lag among variables increases. In this work we propose and develop such generalization. In other words, we propose a generalization of the $\beta$ARMA model of \cite{Rocha2009} by allowing $\tau_t$ to present long-range dependence structure.

\section{The proposed model}

The most widely applied model for time series presenting long-range dependence is the class of autoregressive fractionally integrated moving average (ARFIMA) models, introduced by
\cite{Granger1980} and \cite{Hosking1981} \citep[see also][]{Brockwell1991}.
Recall that a process $\{z_t\}_{t=1}^\infty$ is called an ARFIMA$(p,d,q)$ if it is a weakly stationary solution of {
\[\phi(L)(1-L)^dz_t=\theta(L)\varepsilon_t.\]}
Here $L$ denotes the backward shift operator $L^k(z_t)=z_{t-k}$, for $k\in\{1,2,\dots\}$, {$\varepsilon_t$} is an error term (usually taken as a white noise), $\phi(z)$ and $\theta(z)$ denote the AR and MA polynomials given respectively by
\[\phi(z)=-\sum_{i=0}^p\phi_iz^i,\qquad\theta(z)=\sum_{j=0}^q\theta_jz^j, \quad \forall z\in\C,\]
assumed, as usual, to present no common roots, where $\phi_0=-1$ and $\theta_0=1$. The fractional term $(1-L)^d$ is defined by its binomial expansion, which, in more useful form reads
\begin{align}\label{arf}
(1-L)^{-d}=\sum_{k=0}^\infty\pi_kL^k,\quad \mbox{where}\quad \pi_k=\frac{\Gamma(k+d)}{\Gamma(k+1)\Gamma(d)}=\prod_{j=1}^k \frac{j-1+d}j, \ k\geq1,
\end{align}
and $\pi_0=1$. In this work we shall assume $d\in(-0.5,0.5)$. In this range, it can be shown that if the polynomial $\phi(z)$ does not have roots in the unitary disk $\{z\in\C:|z|=1\}$, then the ARFIMA$(p,d,q)$ is {weakly} stationary.
More details on the theory of ARFIMA processes can be found in \cite{Brockwell1991} and \cite{palma2007}.

In this work we propose a generalization of the $\beta$ARMA model by allowing $\tau_t$ to follow an ARFIMA$(p,d,q)$ process. To motivate the model, {following a similar approach} as \cite{Rocha2009},  assume that, conditionally to $\F_{t-1}$, $\{g(y_t)-\bs{x}_{t-1}'\bs{\beta}\}_{t=1}^\infty$ is a zero-mean stationary ARFIMA$(p,d,q)$ process and write
\begin{align}\label{premodel}
&\phi(L)(1-L)^d\big(g(y_t)-\bs{x}_{t-1}'\bs{\beta}\big)=\theta(L)r_t \nonumber \\
% & \Longrightarrow\quad \phi(L)\big(g(y_t)-\bs{x}_{t-1}'\bs{\beta})=(1-L)^{-d}\theta(L)(r_t) \nonumber\\
& \Longrightarrow\quad g(y_t)=\bs{x}_{t-1}'\bs{\beta}+\sum_{j=1}^p\phi_j\big(g(y_{t-j})-\bs{x}_{t-j-1}'\bs{\beta}\big)+r_t+\sum_{k=1}^\infty c_kr_{t-k},
\end{align}
where we define $\theta_k=0,$ for $k>q$ and $r_t$ is an $\F_{t}$-measurable error term satisfying $\E(r_t|\F_{t-1})=0$, for all $t>1$. {The error term is defined in a recursive fashion in the prediction scale, that is, we consider $r_t=g(y_t)-g(\mu_t)$.}
%\bayer{Pessoal, acho que nao pode ser qualquer $r_t$. Para sair da equacao 4 e chegar na 6 soh se for $r_t=g(y_t)-g(\mu_t)$, certo? Para $r_t=y_t-\mu_t$ nao dah, por exemplo. }
The $c_k$'s in \eqref{premodel}  are the coefficients obtained from Laurent's expansion of $(1-z)^{-d}\theta(z)$, namely
\begin{align}\label{cks}
c_0=1, \quad\mbox{and}\quad  c_k=\sum_{i=0}^{\min\{k,q\}}\theta_i\pi_{k-i}, \ k> 0.
\end{align}
Taking conditional expectation with respect to $\F_{t-1}$ in \eqref{premodel}, noticing that $\E\big(g(y_t)-\bs{x}_{t-1}'\bs{\beta}|\F_{t-1}\big)\approx \tau_t$,
%we obtain the approximate model
%\[\tau_t=\sum_{j=1}^p\phi_j\big(g(y_{t-j})-\bs{x}_{t-j-1}'\bs{\beta}\big)+ \sum_{k=1}^\infty c_kr_{t-k}, \]
upon substituting $\tau_t=g(\mu_t)-\bs{x}_{t-1}'\bs{\beta}$ and adding an intercept $\alpha\in\R$ for $g(\mu_t)$, we arrive at
\begin{align}\label{model}
\eta_t=g(\mu_t)=\alpha + \bs{x}_{t-1}'\bs{\beta}+\sum_{j=1}^p\phi_j\big(g(y_{t-j})-\bs{x}_{t-j-1}'\bs{\beta}\big)+\sum_{k=1}^\infty c_kr_{t-k},
\end{align}
with $c_k$ given in \eqref{cks}. {Observe that the righthand side of \eqref{model} is a real number, hence $\eta_t\in\R$ for all $t$.} Specification \eqref{p1} together with \eqref{model} define the proposed $\beta$ARFIMA$(p,d,q)$ model.

\section{Parameter estimation}

In this section we shall derive the partial maximum likelihood estimator for the parameters in the proposed $\beta$ARFIMA model. Let $\{(y_t,\bs{x}_t^\prime)\}_{t=1}^n$ a sample from a $\beta$ARFIMA$(p,d,q)$ model. Let us denote the $(p+q+l+3)$-dimensional parameter vector by  $\bs\gamma'=(\nu, d,\alpha, \bs\beta', \bs\phi',\bs\theta')$  and let $\Omega\subseteq (0,\infty)\times (-0.5,0.5)\times \R^{p+q+l+1}$ be the parameter space.
By letting
\begin{align*}
\ell_t(\mu_t,\nu)&=\log\big(f(y_t;\mu_t,\nu|\F_{t-1})\big)\\
&=\log\big(\ga(\nu)\big)-\log\big(\ga(\mu_t\nu)\big)-\log\big(\ga\big(\nu(1-\mu_t)\big)\big)+\\
&\qquad\quad +(\mu_t\nu-1)\log(y_t)+\big(\nu(1-\mu_t)-1\big)\log(1-y_t),\\
\end{align*}
the partial log-likelihood function is given by
\begin{align}\label{E:loglik}
\ell(\bs\gamma)=\sum_{t=1}^{n}\ell_t(\mu_t,\nu)
\end{align}
and hence, the partial maximum likelihood estimator of $\bs\gamma$ is defined as
\begin{align}\label{MLE}
\widehat{\bs\gamma}=\underset{\bs\gamma\in\Omega}{\mathrm{argmax}}\big\{\ell(\bs\gamma)\big\}.
\end{align}
In the next section we shall derive the score vector related to the maximization problem \eqref{MLE}.

\subsection{Partial score vector}\label{s:score}

To derive the partial score vector we shall need to obtain the derivative of the log-likelihood $\ell(\bs\gamma)$ given in \eqref{E:loglik} with respect to each coordinate $\gamma_j$ of the parameter $\bs\gamma'=(\nu, d,\alpha, \bs\beta', \bs\phi',\bs\theta')$.
Let $\psi:(0,\infty)\rightarrow\R$ be the digamma function defined as $\psi(z)=\frac{d}{dz}\log\big(\Gamma(z)\big)$. The derivative of $\ell(\bs\gamma)$ with respect to $\nu$ can be easily obtained as
\begin{align}\label{e:dldnu}
\frac{\partial \ell(\bs\gamma)}{\partial \nu}=\sum_{t=1}^n \big[\mu_t(y_t^\ast-\mu_t^\ast)+\log(1-y_t)-\psi\big((1-\mu_t)\nu\big)+\psi(\nu)\big],
\end{align}
where
\begin{align*}%\label{stars}
y_t^ {\ast} = \log \bigg(\frac{y_t}{1-y_t}\bigg) \quad \mbox{and}\quad \mu_t^ {\ast} = \psi(\mu_t\nu)-\psi\big((1-\mu_t)\nu\big).
\end{align*}
The derivative of the log-likelihood $\ell(\bs\gamma)$ with respect to the remaining parameters $\gamma_j \neq \nu$ can be computed by the general differentiation rule
\begin{align}\label{genrule_a}
 \frac{\partial \ell(\bs\gamma)}{\partial \gamma_j} =
 \sum_{t=1}^{n} \frac{\partial \ell_t(\mu_t,\nu)}{\partial \mu_t}
 \frac{d \mu_t}{d \eta_t} \frac{\partial \eta_t}{\partial \gamma_j}.
\end{align}
Observe that
\begin{align}\label{e:dldmu}
\frac{\partial \ell_t(\mu_t,\nu)}{\partial \mu_t} = \nu
\bigg[
\log \bigg(\frac{y_t}{1-y_t}\bigg) - \psi (\mu_t \nu) + \psi \big( \nu(1-\mu_t) \big)
\bigg]=\nu(y_t^\ast-\mu_t^\ast).
\end{align}
Since $\eta_t=g(\mu_t)$, it also follows that  $\displaystyle{\frac{d \mu_t}{d \eta_t} = \frac{1}{g'(\mu_t)}}$. Substituting these results in \eqref{genrule_a}, we obtain
\begin{align}\label{genrule}
 \frac{\partial \ell(\bs\gamma)}{\partial \gamma_j} =
 \sum_{t=1}^{n} \frac{\nu(y_t^\ast-\mu_t^\ast)}{g^\prime(\mu_t)}\frac{\partial \eta_t}{\partial \gamma_j}\,.
\end{align}
Hence, the task of computing the derivatives of $\ell(\bs\gamma)$ greatly simplifies to determining the derivatives $\displaystyle{\frac{\partial \eta_t}{\partial \gamma_j}}$, for each coordinate $\gamma_j \neq \nu$ of the parameter vector $\bs\gamma$.  To obtain the derivative with respect to $d$, {recall that $r_t=g(y_t)-g(\mu_t)$ and notice that}
\[\df{\eta_t}{d} =\sum_{k=1}^{\infty}\bigg[r_{t-k}\df{c_k}{d}+c_k\df{r_{t-k}}{d}\bigg]=\sum_{k=1}^{\infty}\bigg[r_{t-k}\df{c_k}{d}-c_k\df{\eta_{t-k}}{d}\bigg].\]
For $\pi_k$ given in \eqref{arf}, by using the identity $\Gamma^\prime(x)=\Gamma(x)\psi(x)$, it follows that
\begin{align*}
\df{\pi_{m}}{d}&=\frac{1}{\Gamma(m+1)\Gamma(d)^2}\left[\Gamma(d)\df{\Gamma(d+m)}{d}-\Gamma(d+m)\df{\Gamma(d)}{d}\right]\\
&=\pi_{m}\big[\psi(d+m)-\psi(d)\big].
\end{align*}

Hence
%\[\df{c_k}{d}=\sum_{i=0}^{\min\{k,q\}} \theta_i\df{\pi_{k-i}}{d}=\sum_{i=0}^{\min\{k,q\}} \theta_i\pi_{k-i}\big[\psi(d+k-i)-\psi(d)\big]\]
%and
\begin{align*}
\df{\eta_t}{d} = \sum_{k=1}^\infty\bigg(r_{t-k}\sum_{i=0}^{\min\{k,q\}}\theta_i\pi_{k-i}\big[\psi(d+k-i)-\psi(d)\big]-c_k\df{\eta_{t-k}}{d}\bigg).
%-\sum_{k=1}^\infty\bigg(&r_{t-k}\sum_{i=0}^{k-1}\sum_{j=0}^{k-i}\phi_j\bigg[\frac{\Gamma(d+k-i-j)c_i}{\Gamma(k-i-j+1)\Gamma(d)}\big[\psi(d+k-i-j)-\psi(d)\big]+\pi_{k-i-j}\df{c_i}{d}\bigg]\\
%&+c_k\df{\eta_{t-k}}{d}\bigg).
\end{align*}
Differentiation with respect to $\alpha$ yields
\begin{align*}
\frac{\partial \eta_t}{\partial \alpha}= 1 + \sum_{k=1}^{\infty} c_k\frac{\partial r_{t-k}}{\partial \alpha} =1 - \sum_{k=1}^{\infty} c_k\frac{\partial \eta_{t-k}}{\partial \alpha}.
\end{align*}
%so that
%\begin{align*}
% \frac{\partial \ell(\bs\gamma)}{\partial \alpha}
% =\sum \limits_{t=1}^{n}\frac{\nu(y_t^\ast-\mu_t^\ast)} {g^\prime(\mu_t)}
%\bigg[1- \sum_{k=1}^{\infty} c_k\frac{\partial \eta_{t-k}}{\partial \alpha}\bigg]\, .
%\end{align*}
Regarding parameter $\beta_s$, for $s\in\{1,\dots,l\}$, we have
\[\df{\eta_t}{\beta_s}=x_{t-1}(s)-\sum_{j=1}^p \phi_jx_{t-j-1}(s)-\sum_{k=1}^{\infty} c_k\df{\eta_{t-k}}{\beta_s},\]
where $x_{m}(s)$ denotes the $(s)$-th element of $\bs{x}_{m}$.
%\[\frac{\partial \ell(\bs\gamma)}{\partial \beta_s}= \sum_{t=1}^{n} \frac{\nu(y_t^\ast-\mu_t^\ast)}{g^\prime(\mu_t)}\bigg[x_{t-1}(s)-\sum_{k=1}^p \phi_kx_{t-k-1}(s)-\sum_{k=1}^{\infty} c_k\df{\eta_{t-k}}{\beta_s}\bigg].\]
%As for the log-likelihood derivative with respect to the autoregressive parameters, $\phi_s$, for $s\in\{1,\dots,p\}$, we have
The log-likelihood derivative with respect to $\phi_s$, for $s\in\{1,\dots,p\}$, is given by
\[\df{\eta_t}{\phi_s}=g(y_{t-s})-\bs{x}_{t-s-1}^\prime\bs\beta-\sum_{k=1}^\infty c_k\df{\eta_{t-k}}{\phi_s}.\]
%hence
%\begin{align*}
%\frac{\partial \ell(\bs\gamma)}{\partial \phi_s} =
%\sum_{t=1}^{n} \frac{\nu(y_t^\ast-\mu_t^\ast)}{g^\prime(\mu_t)}\bigg[g(y_{t-s})-\bs{x}_{t-s-1}^\prime\bs\beta-\sum_{k=1}^\infty c_k\df{\eta_{t-k}}{\phi_s}\bigg].
% \end{align*}
For the log-likelihood derivative with respect to $\theta_s$, with $s=1,\dots,q$, we have
\begin{align*}
\df{\eta_t}{\theta_s}=\sum_{k=1}^\infty\bigg[r_{t-k}\df{c_k}{\theta_s}+c_k\df{r_{t-k}}{\theta_s}\bigg].
\end{align*}
Now, differentiating \eqref{cks} with respect to $\theta_s$, we obtain $\displaystyle{ \df{c_k}{\theta_s}}=\pi_{k-s}I(k\geq s)$ so that, from \eqref{genrule},
\begin{align*}
 \frac{\partial \ell(\bs\gamma)}{\partial \theta_s} = \sum \limits _ {t=1} ^ {n}  \frac{\nu(y_t^\ast - \mu_t^\ast)}{g^\prime(\mu_t)}
 \bigg[\sum_{k=s}^\infty \pi_{k-s}r_{t-k}-\sum_{k=1}^\infty c_k\df{\eta_{t-k}}{\theta_s}\bigg].
\end{align*}
Finally, let $M$ be the $n\times l$ matrix whose $(t,s)$-th element is given by
\[M_{t,s}=\df{\eta_t}{\beta_s}=x_{t-1}(s)-\sum_{k=1}^p \phi_kx_{t-k-1}(s)-\sum_{k=1}^{\infty} c_k\df{\eta_{t-k}}{\beta_s},\]
$P$ be the $n\times p$ matrix whose $(t,s)$-th element is given by
\[P_{t,s}=\df{\eta_t}{\phi_s}=g(y_{t-s})-\bs{x}_{t-s-1}^\prime\bs\beta-\sum_{k=1}^\infty c_k\df{\eta_{t-k}}{\phi_s}\]
and $Q$ be the $n\times q$ matrix whose $(t,s)$-th element is given by
\[Q_{t,s}=\df{\eta_t}{\theta_s}=\sum_{k=s}^\infty \pi_{k-s}r_{t-k}-\sum_{k=1}^\infty c_k\df{\eta_{t-k}}{\theta_s}.\]
Let $\bs{\mathcal{Y}}=(y_1^\ast - \mu_1^\ast,\dots,y_n^\ast - \mu_n^\ast)^\prime$, $\bs{\texttt{d}}=\big(\df{\eta_1}{d},\dots,\df{\eta_n}{d}\big)'$, $\bs{a}=\big(\df{\eta_1}{\alpha},\dots,\df{\eta_n}{\alpha}\big)'$ and $T=\mathrm{diag}\big\{g'(\mu_1)^{-1},\dots,g'(\mu_n)^{-1}\big\}$.
Then, the score vector can be written in matrix form as
\[U(\bs\gamma)=\big(U_\nu(\bs\gamma), U_d(\bs\gamma),U_\alpha(\bs\gamma),U_\beta(\bs\gamma)^\prime,U_\phi(\bs\gamma)^\prime,U_\theta(\bs\gamma)^\prime\big)^\prime\in\R^{p+q+l+3},\]
where
\begin{align*}
U_d(\bs\gamma)&=\nu\bs{\texttt{d}}'T\bs{\mathcal{Y}},\qquad\qquad
U_\alpha(\bs\gamma)=\nu \bs{a}'T\bs{\mathcal{Y}}, \qquad\qquad
U_{\beta}(\bs\gamma)=\nu M'T\bs{\mathcal{Y}}, \\
U_\phi(\bs\gamma)&=\nu P'T\bs{\mathcal{Y}}, \qquad\qquad
U_\theta(\bs\gamma)=\nu Q'T\bs{\mathcal{Y}},\\
\mbox{and}\quad U_\nu(\bs\gamma)&=\sum_{t=1}^n \big[\mu_t(y_t^\ast-\mu_t^\ast)+\log(1-y_t)-\psi\big((1-\mu_t)\nu\big)+\psi(\nu)\big].
\end{align*}

The conditional maximum likelihood is obtained by
numerically
solving the nonlinear system $U(\bs\gamma)=\bs{0}$, where $\bs{0}$ denotes the null vector in $\R^{p+q+l+3}$. %The system has no explicit solution, but an approximate solution can be found by numerical methods.

\subsection{Conditional information matrix}

In this section we derive the Fisher conditional information matrix, which will be useful later on deriving the asymptotic properties of the partial maximum likelihood estimator for the proposed model.
For $i,j\in\{1,\dots,p+q+l+3\}$, it can be shown that
\begin{align*}
\frac{\partial^2\ell(\bs\gamma)}{\partial \gamma_i \partial \gamma_j} &= \sum_{t=1}^{n}\frac{\partial}{\partial \mu_t}
\left( \frac{\partial \ell_t(\mu_t,\nu)}{\partial \mu_t}\frac{d \mu_t}{d \eta_t} \frac{\partial \eta_t}{\partial \gamma_j}\right)
\frac{d \mu_t}{d \eta_t} \frac{\partial \eta_t}{\partial \gamma_i} \\
&= \sum_{t=1}^{n} \left[ \frac{\partial^2 \ell_t(\mu_t,\nu)}{\partial \mu_t^2}\frac{d \mu_t}{d \eta_t} \frac{\partial \eta_t}{\partial \gamma_j}
+ \frac{\partial \ell_t(\mu_t,\nu)}{\partial \mu_t}\frac{\partial}{\partial \mu_t}\left(\frac{d \mu_t}{d \eta_t} \frac{\partial \eta_t}{\partial \gamma_j} \right) \right]
\frac{d \mu_t}{d \eta_t} \frac{\partial \eta_t}{\partial \gamma_i}\,.
\end{align*}
Under the regularity conditions presented in Section \ref{asymptotictheory}, $\E\big(\partial \ell_t(\mu_t,\nu)/\partial \mu_t \big| \F_{t-1}\big)=0$, so that
\begin{align*}
\E\left( \left. \frac{\partial^2\ell(\bs\gamma)}{\partial \gamma_i \partial \gamma_j}  \right| \F_{t-1} \right)
&= \sum_{t=1}^{n} \E\left( \left. \frac{\partial^2 \ell_t(\mu_t,\nu)}{\partial \mu_t^2} \right| \F_{t-1} \right)
\left(\frac{d \mu_t}{d \eta_t} \right)^2
\frac{\partial \eta_t}{\partial \gamma_j}
\frac{\partial \eta_t}{\partial \gamma_i}.
\end{align*}
Deriving \eqref{e:dldmu} with respect to $\mu_t$ twice, we obtain
\begin{align*}
\frac{\partial^2 \ell_t(\mu_t,\nu)}{\partial \mu_t^2} = -\nu^2
\left\{
\psi' (\mu_t \nu) +
\psi'
\left[
(1-\mu_t)\nu
\right]
\right\}.
\end{align*}
Set $w_t=\nu^2 \big\{ \psi' (\mu_t \nu) + \psi' \big[ (1-\mu_t)\nu \big] \big\}$. Since $\mu_t$ is $\F_{t-1}$-measurable, we obtain %a general expression for the necessary expectations is given by
\begin{align*}%\label{gen2der}
\E\left( \left. \frac{\partial^2\ell(\bs\gamma)}{\partial \gamma_i \partial \gamma_j}  \right| \F_{t-1} \right)
= - \sum_{t=1}^{n} \frac{w_t}{g'(\mu_t)^2} \frac{\partial \eta_t}{\partial \gamma_j} \frac{\partial \eta_t}{\partial \gamma_i}.
\end{align*}
%where $w_t=\nu^2 \big\{ \psi' (\mu_t \nu) + \psi' \big[ (1-\mu_t)\nu \big] \big\} $. %The particular format \eqref{gen2der} takes is appealing due to the fact that the derivatives $\df{\eta_t}{\gamma_j}$ have already been obtained in the Section~\ref{s:score}.
Direct differentiation of $\frac{\partial \ell(\bs\gamma)}{\partial \nu}$ with respect to $\gamma_i$, yields
\begin{align*}
\frac{\partial^2 \ell(\bs\gamma)}{\partial \nu \partial \gamma_i} = \sum \limits _ {t=1} ^ {n}
\bigg[ (y_t^\ast - \mu_t^\ast) - \nu \frac{\partial \mu_t^\ast}{\partial \nu} \bigg]
\frac{1}{g'(\mu_t)}\frac{\partial \eta_t}{\partial \gamma_i},
\end{align*}
where ${\partial \mu_t^\ast}/{\partial \nu} =
\psi' (\mu_t \nu) \mu_t
-\psi'
\left[
(1-\mu_t)\nu
\right] (1-\mu_t)$.
Under some regularity conditions (see Section \ref{asymptotictheory}), we have $\E(y^\ast_t | \F_{t-1})=\mu_t^\ast$, and thus
\begin{align*}
\E\left( \left. \frac{\partial^2 \ell(\bs\gamma)}{\partial \nu \partial \gamma_i}  \right| \F_{t-1} \right)
&= - \sum_{t=1}^{n}
\frac{v_t}{g'(\mu_t)}
\frac{\partial \eta_t}{\partial \gamma_i},
\end{align*}
where $ v_t = \nu \frac{\partial \mu_t^\ast}{\partial \nu} = \nu \big\{\psi' (\mu_t \nu) \mu_t -\psi'\big[(1-\mu_t)\nu \big] (1-\mu_t)\big\}$. The expected value of the second derivative of $\ell(\bs\gamma)$ with respect to $\nu$ is given by
\begin{align*}
\E\left( \left. \frac{\partial^2 \ell(\bs\gamma)}{\partial \nu^2}  \right| \F_{t-1} \right) =
- \sum_{t=1}^{n}S_t,
\end{align*}
where $S_t=\psi' (\mu_t\nu)\mu_t^2+\psi'\big[(1-\mu_t)\nu\big](1-\mu_t)^2-\psi'(\nu)$.

Finally, let $\bs{v}=(v_1,\dots,v_n)'$,  $W = \mathrm{diag}\Big\{\frac{w_1}{g'(\mu_1)^2},\dots,\frac{w_n}{g'(\mu_n)^2}\Big\}$,
%$R=\bigg(\df{\eta_1}{\alpha},\dots,\df{\eta_n}{\alpha}\bigg)'$,
and \linebreak $S=\mathrm{diag}\{S_1,\dots,S_n\}$, in matrix form we have
%\pagebreak
\begin{align*}
\begin{array}{ccc}
E\left( \left. \frac{\partial^2 \ell(\bs\gamma)}{\partial \nu^2}  \right| \F_{t-1} \right)=-\tr(S),                               \phantom{XXXX}&
E\left( \left. \frac{\partial^2 \ell(\bs\gamma)}{\partial \nu \partial d }  \right| \F_{t-1} \right)=-\bs{v}'T\bs{\texttt{d}},             \vspace{.3cm}\\
E\left( \left. \frac{\partial^2 \ell(\bs\gamma)}{\partial \nu \partial \alpha }  \right| \F_{t-1} \right)=-\bs{v}'T\bs{a},                  \phantom{XXXX}&
E\left( \left. \frac{\partial^2 \ell(\bs\gamma)}{\partial \bs\beta \partial \nu  }  \right| \F_{t-1} \right)=-M'T\bs{v},               \vspace{.3cm}\\
E\left( \left. \frac{\partial^2 \ell(\bs\gamma)}{\partial \bs\phi \partial \nu }  \right| \F_{t-1} \right)=-P'T\bs{v},                 \phantom{XXXX}&
E\left( \left. \frac{\partial^2 \ell(\bs\gamma)}{\partial \bs\theta \partial \nu }  \right| \F_{t-1} \right)=-Q'T\bs{v},              \vspace{.3cm}\\
E\left( \left. \frac{\partial^2 \ell(\bs\gamma)}{\partial d^2}  \right| \F_{t-1} \right)=-\bs{\texttt{d}}'W\bs{\texttt{d}},              \phantom{XXXX}&
E\left( \left. \frac{\partial^2 \ell(\bs\gamma)}{\partial d \partial \alpha}  \right| \F_{t-1} \right)=-\bs{\texttt{d}}'W\bs{a},  \vspace{.3cm}\\
E\left( \left. \frac{\partial^2 \ell(\bs\gamma)}{\partial \bs\beta \partial d  }  \right| \F_{t-1} \right)=-M'W\bs{\texttt{d}},       \phantom{XXXX}&
E\left( \left. \frac{\partial^2 \ell(\bs\gamma)}{\partial \bs\phi \partial d  }  \right| \F_{t-1} \right)=-P'W\bs{\texttt{d}},        \vspace{.3cm}\\
E\left( \left. \frac{\partial^2 \ell(\bs\gamma)}{\partial \bs\theta \partial d  }  \right| \F_{t-1} \right)=-Q'W\bs{\texttt{d}},      \phantom{XXXX}&
E\left( \left. \frac{\partial^2 \ell(\bs\gamma)}{\partial \alpha^2}  \right| \F_{t-1} \right)=-\tr(W),                           % \vspace{.3cm}\\
\end{array}
\end{align*}
\begin{align*}
\begin{array}{ccc}
E\left( \left. \frac{\partial^2 \ell(\bs\gamma)}{\partial \bs\beta \partial \alpha}  \right| \F_{t-1} \right)=-M'W\bs{a},     \phantom{XXXX}&
E\left( \left. \frac{\partial^2 \ell(\bs\gamma)}{\partial \bs\phi \partial \alpha}  \right| \F_{t-1} \right)=-\bs{a}'WP,      \vspace{.3cm}\\
E\left( \left. \frac{\partial^2 \ell(\bs\gamma)}{\partial \bs\theta \partial \alpha}  \right| \F_{t-1} \right)=-Q'W\bs{a},    \phantom{XXXX}&
E\left( \left. \frac{\partial^2 \ell(\bs\gamma)}{\partial \bs\beta \partial \bs\beta'}  \right| \F_{t-1} \right)=-M'WM,           \vspace{.3cm}\\
E\left( \left. \frac{\partial^2 \ell(\bs\gamma)}{\partial \bs\beta \partial \bs\phi'}  \right| \F_{t-1} \right)=-M'WP,            \phantom{XXXX}&
E\left( \left. \frac{\partial^2 \ell(\bs\gamma)}{\partial \bs\beta \partial \bs\theta'}  \right| \F_{t-1} \right)=-M'WQ,          \vspace{.3cm}\\
E\left( \left. \frac{\partial^2 \ell(\bs\gamma)}{\partial \bs\phi \partial \bs\phi'}  \right| \F_{t-1} \right)=-P'WP,             \phantom{XXXX}&
E\left( \left. \frac{\partial^2 \ell(\bs\gamma)}{\partial \bs\phi \partial \bs\theta'}  \right| \F_{t-1} \right)=-P'WQ,           \vspace{.3cm}\\
E\left( \left. \frac{\partial^2 \ell(\bs\gamma)}{\partial \bs\theta \partial \bs\theta'}  \right| \F_{t-1} \right)=-Q'WQ.         \phantom{XXXX}&
\end{array}
\end{align*}
Now, for $i,j\in\{1,\dots,p+q+l+3\}$ let
\begin{align*}%\label{eq:G}
G_{\bs\gamma_i,\bs{\gamma}_j}(\bs{\gamma})=-E\left( \left. \frac{\partial^2 \ell(\bs\gamma)}{\partial \bs\gamma_i \partial \bs\gamma_j'}  \right| \F_{t-1} \right), \quad \mbox{ for $i\leq j$,}
 \end{align*}
and $G_{\bs\gamma_j,\bs{\gamma}_i}(\bs\gamma)=G_{\bs\gamma_i,\bs{\gamma}_j}(\bs{\gamma})'$, for $i>j$.
Hence, the Fisher information matrix is the matrix $G_n(\bs{\gamma})$ whose $(i,j)$-th element is $G_{\bs\gamma_i,\bs{\gamma}_j}(\bs{\gamma})$.

\section{Asymptotic theory and hypothesis testing}\label{asymptotictheory}

Rigorous asymptotic theory for the maximum likelihood estimator in the context of generalized linear models for canonical link functions was first developed in \cite{Haberman1977a} and \cite{Nordberg1980}. For non-canonical links, the work of \cite{Fahrmeir1985} was a pioneer, setting grounds for latter development of the theory. The work of \cite{Wong1986} develops the theory of partial likelihood in the context of non-Gaussian and non-stationary time series.  For GARMA-like models \citep{Benjamin1998, Rocha2009}, a general theory for PMLE is presented in the works of \cite{Fokianos1998,Fokianos2004}, from which we build up upon, and (at some extent) \cite{Li1991}. {We remark that, since the $\beta$ARMA model can be viewed as a special case of our model (when $d=0$ and the covariates are all non-random) the asymptotic theory presented here completes the one presented in \cite{Rocha2009} and \cite{Rocha2017}.}
Let $\{(y_t,\bs{x}_t)\}_{t=1}^n$ be a sample from a $\beta$ARFIMA$(p,d,q)$ model specified by \eqref{p1} and \eqref{model}. Let  $U(\bs\gamma)$ denote the partial score vector based on the sample and $\widehat{\bs\gamma}$ denote a solution of $U(\bs\gamma)=\bs0$. {Also, for $j\in\N$ let
$\bs{h}(t,j)=g(y_{t-j})-P_{\{\bs{x}_1',\dots,\bs{x}_{t-j-1}'\}}\big(g(y_j)\big)$ where $P_{\{\bs{x}_1',\dots,\bs{x}_{t-j-1}'\}}\big(g(y_j)\big)$ denotes the projection of $g(y_j)$ into the space generated by $\bs{x}_1',\dots,\bs{x}_{t-j-1}'$} and
\[\bs Z_t=\big(1,\bs{x}_{t-1}',\bs{h}(t,1),\dots,\bs{h}(t,p), r_{t-1},r_{t-2},\dots\big)'.\]
To calculate the PMLE $\widehat \gamma_n$ from a sample, we approximate the derivatives by truncating the infinite moving average representation \eqref{model} to a point  $m$, initialize $r_{t}=0$ and $\mu_t=0$ for $t\leq p$ and calculate $\mu_t$ and $r_t$ for $t>p$ recursively from the data through \eqref{model}.

The required regularity conditions for the asymptotic existence/uniqueness, the consistency and the asymptotic normality of the partial likelihood estimator for the $\beta$ARFIMA models are fundamentally the same as in \cite{Fokianos1998,Fokianos2004}.

\subsubsection*{Assumptions}

\begin{enumerate}[(A)]
\item The inverse link function $g^{-1}$ is of class $\mathcal{C}^2$ and {satisfies $\big|\partial g^{-1}(x)/\partial x\big|\neq0$}, for all $x\in\R$.
\item The parametric space $\Omega$ is an open set in $\R^{l+q+p+3}$ and the {true parameter $\bs\gamma_0$ lies in $\Omega$}.
\item For each $t$, the covariate vector $\bs{Z}_{t}$ almost surely belongs to a compact set $\Upsilon\subset\Omega$ and there exists $n_0\in \N$ such that, for all $n>n_0$, $P(\sum_{t=1}^n\bs{Z}_{t}\bs{Z}_{t}'>0)=1$. Additionally, assume that $g^{-1}(\eta_t)$ is almost surely well-defined for all $\bs{Z}_{t}\in\Upsilon$ and {$\bs\gamma \in {\Omega}$}.
\item There exists a probability measure $\lambda$ in $\Omega$ such that $\int_{\Omega}\bs{z}\bs{z}'\lambda(d\bs{z})$ is positive definite and such that the weak convergence
    \[\frac1{n}\sum_{t=1}^nI(\bs{Z}_{t-1}\in A)\underset{n\rightarrow\infty}\longrightarrow \lambda(A),\]
holds for all $\lambda$-continuity sets $A\subset \Omega$ under \eqref{model} with {$\bs\gamma=\bs\gamma_0$}.
\end{enumerate}

Assumptions A and B guarantee that $\partial^2 \ell(\bs\gamma)/\partial\bs\gamma'\partial\bs\gamma$ is a continuous function of $\bs\gamma$, while Assumptions B and C imply that, for all sufficiently large $n$, the conditional information matrix is positive definite. Assumptions A, B, and C also assure that the model is well defined. The compactness assumption in C is mathematically convenient. It can, however, be replaced by the requirement that there exists an increasing sequence of compact sets, $\{\Upsilon_n\}_{n=1}^\infty$, say, such that, for sufficiently large $n$, $\bs{x}_{n}\in\Upsilon_n$ with high probability. A probability measure satisfying the weak convergence in Assumption D also satisfies
\[\frac1n\sum_{t=1}^nf(\bs Z_{t-1})\overset{P}{\underset{n\rightarrow\infty}\longrightarrow}\int_\Upsilon f(\bs z)\lambda(d\bs z),\]
for all bounded and continuous function $f:\Upsilon\rightarrow\R$,
and implies the weak convergence of the conditional information matrix to a (non-random) positive definite matrix in the sense that there exists a matrix, which we denote by $G(\bs\gamma)$, such that
\[\frac{G_n(\bs\gamma)}n\underset{n\rightarrow\infty}{\longrightarrow} G(\bs\gamma),\quad \forall \bs\gamma\in\Omega.\]
Observe that assumption D implies that $G(\bs\gamma_0)$ is positive definite and its inverse exists. Conditions C and D also imply conditions C and D in \cite{Fahrmeir1985}, which, in turn, imply the asymptotic existence of a sequence of solutions for $U(\bs\gamma)=\bs0$. See also the discussion on \cite{Fahrmeir1985} and \cite{Fokianos1998,Fokianos2004}.

\begin{theorem}\label{an}
Under the assumptions A-D, the probability that a locally unique maximum partial likelihood estimator exists in a neighborhood of $\bs\gamma_0$ tends to one. Furthermore, the estimator is consistent
\[\widehat{\bs\gamma}\overset{P}{\underset{n\rightarrow\infty}\longrightarrow} \bs\gamma_0\]
and asymptotically normal
\[\sqrt{n}(\widehat{\bs\gamma}-\bs\gamma_0)\overset{d}{\underset{n\rightarrow\infty}\longrightarrow} N_{p+q+l+3}\big(\bs0,G(\bs\gamma_0)^{-1}\big).\]
%and satisfy
%\[\sqrt{n}\big(\widehat{\bs\gamma}-\bs\gamma_0-G^{-1}(\bs\gamma_0)U(\bs\gamma_0)\big) \overset{P}{\underset{n\rightarrow\infty}\longrightarrow} \bs0.\]
\end{theorem}

\proof {The proof follows the same lines as the proof of Theorem 3.1 in \cite{Fokianos1998} by the $\F_{t}$-measurability of $r_t$ and since, under the hypothesis, $\bs{h}(t,j)\rightarrow\bs{x}_{t-j-1}'\bs\beta$, almost surely. The key point is to show that the score vector is a zero mean square integrable martingale sequence with respect to an adequate filtration. Consider the filtration $\{\F_{t}, t\in\Z\}$ where $\F_{t}=\sigma\{Y_{t}, Y_{t-1},\dots, \bs {x}_{t},\bs x_{t-1},\dots\}$ and let $\{U_t(\bs\gamma)\}_{t\in\Z}$ denote the partial score process given by
\[U_t(\bs\gamma):=\left(\frac{\partial \ell_t(\bs\gamma)}{\partial \gamma_1},\dots,\frac{\partial \ell_t(\bs\gamma)}{\partial \gamma_{l+q+p+3}}\right)^\prime.\]
First observe that the $\F_{t-1}$-measurability of $\mu_t$ and assumptions $A$ and $B$ imply that $\{U_t(\bs\gamma)\}_{t\in\Z}$ is integrable and adapted to the filtration $\{\F_{t}, t\in\Z\}$. For $\gamma_j\neq\nu$ it is straightforward to show that, under assumptions A and B, $\E\big(y_t^\ast|\F_{t-1}\big)=\mu_t^\ast$, hence, by \eqref{e:dldmu}, we conclude that $\E\big(\partial \ell_t(\mu_t,\nu)/\partial\mu_t \big|\F_{t-1}\big)=0$. When $\gamma_j=\nu$, the result follows from \eqref{e:dldnu} since $\E\big(\log(1-y_t)|\F_{t-1}\big)=\psi\big((1-\mu_t)\nu\big)-\psi(\nu)$. The rest of the proof follows the same idea as Theorem 3.1 in \cite{Fokianos1998}, in view of Theorem 1 in \cite{Fokianos2004} and \cite{Kedem2002}.}\hfill\qed
\begin{remark}
In view of the work of \cite{Wong1986}, one could also, in principle, obtain similar large sample results for the partial likelihood considering non-stationary ARFIMA processes in \eqref{model}, under somewhat more stringent conditions.
\end{remark}
\begin{remark}
It is widely known that, under long-range dependence, convergence rates of central limit type theorems are usually slower than the traditional $\sqrt{n}$. Observe, however, that a time series $\{y_t\}$ following a $\beta$ARFIMA model is a sequence of conditionally independent, but not identically distributed random variables. In the presence of time dependent covariates, it is, in fact, non-stationary. The long-range dependence is connected to $g(\mu_t)$, which does not influence the convergence rate {of the corresponding parameters} in Theorem \ref{an}.
\end{remark}

 Let $T:\R^{p+q+l+3}\rightarrow\R^k$, for $k<p+q+l+3$, be a vector valued transformation such that its {Jacobian} $\bs J(\bs\gamma)$ exists, is of full rank $k$ and, as a function of $\bs\gamma$, is continuous in an open subset of $\Omega$. We shall consider composite hypothesis of the form
\begin{align}\label{HT}
\mathcal{H}_0:T(\bs\gamma)=\bs 0 \ \ \mbox{vs.} \ \ \mathcal{H}_1:T(\bs\gamma)\neq \bs 0.
\end{align}
There are several ways to test the restriction \eqref{HT}. Let $\widehat{\bs\gamma}$ be the {unrestricted} PMLE of $\bs\gamma$
%considering the full model
and let $\tilde{\bs\gamma}$ be the PMLE under $\mathcal{H}_0$ in \eqref{HT}.
The partial log-likelihood ratio statistic is given by
\begin{align*}%\label{lpl}
LR=2\big[\ell(\widehat{\bs\gamma})-\ell(\tilde{\bs\gamma})\big].
\end{align*}
The traditional Wald's statistic reads
\begin{align*}
W=nT(\widehat{\bs\gamma})'\big[\bs{J}(\widehat{\bs\gamma})'G^{-1}(\widehat{\bs\gamma})\bs{J}(\widehat{\bs\gamma})\big]^{-1}T(\widehat{\bs\gamma}),
\end{align*}
while the Rao's score statistic is given by
\begin{align*}%\label{SS}
S=\frac1nU(\tilde{\bs\gamma})'G^{-1}(\tilde{\bs\gamma})U(\tilde{\bs\gamma}).
\end{align*}
{Next theorem shows that the asymptotic distribution of the test statistics $LR$, $W$, and $S$ are analogous as their counterparts under independence. The proof is completely analogous to the independent case \citep[see also Theorem 2 in][]{Fokianos2004}.
\begin{theorem}
Under assumptions A-D and under the null hypothesis in \eqref{HT}, the test statistics $LR$, $W$, and $S$ defined above are asymptotically distributed as chi-square with $k$ degrees of freedom.
\end{theorem}}

The square root of the traditional Wald's statistic (often called $z$ statistics)
is particularly convenient to test individual parameters \citep{Pawitan2001}.
Considering the hypothesis
$\mathcal{H}_0:\gamma_j= \gamma_0$ vs. $\mathcal{H}_1:\gamma_j \neq \gamma_0$,
the $z$ statistic is given by
\begin{align}\label{W}
z= \frac{\widehat{\gamma}_j - \gamma_0}{\text{se}(\widehat{\gamma}_j)},
\end{align}
where $\text{se}(\widehat{\gamma}_j)$ is the square root of the $j$-th diagonal element of $G_n(\widehat{\bs \gamma})^{-1}$.
Under $\mathcal{H}_0$, the limiting distribution of $z$ is standard normal.

As an example the transformation $T(\bs\gamma)=d$ coupled with any of the above test statistics can be used to test
\[\mathcal{H}_0: d=0 \ \ \mbox{vs.} \ \ \mathcal{H}_1: d\neq0, \]
which is equivalent to test the presence of long-range dependence in the systematic component of the model. In other words, it can be applied to decide whether a $\beta$ARFIMA or a $\beta$ARMA is suitable to the data, with rejection of the null hypothesis favouring the $\beta$ARFIMA model.

\section{Diagnostic and prediction}

Diagnostics in the context of  $\beta$ARFIMA models follow the usual procedures of GLM theory with some adaptations. For a general goodness of fit testing we consider the so-called deviance statistic.
The deviance $D$ is defined as twice the difference between the conditional log-likelihood of the saturated model (for which $\tilde{\mu}_t=y_t$, i.e., a model with as many parameters as observations) and the fitted model, that is
\begin{align*}
D=2\big(\, \tilde{\ell} - \widehat{\ell}\, \big),\quad\mbox{where}\quad \widehat{\ell}  =  \sum\limits_{t=1}^{n}\ell_t(\widehat{\mu}_t,\widehat{\nu})\ \mbox{ and }\ \tilde{\ell}  =  \sum\limits_{t=1}^{n}\ell_t(y_t,\widehat{\nu}).
\end{align*}
%where $\widehat{\ell}  =  \sum\limits_{t=1}^{n}\ell_t(\widehat{\mu}_t,\widehat{\nu})$ and $\tilde{\ell}  =  \sum\limits_{t=1}^{n}\ell_t(y_t,\widehat{\nu})$.
If the fitted model is correct,
the test statistic $D$ is approximately distributed as chi-squared with $n-(p+q+l+3)$ degrees of freedom \citep{Benjamin2003, Kedem2002, Fokianos2004}.
%See also the discussion in \cite{Fokianos2004}.
%A good rule of thumb in deviance analysis is to divide the deviance by its degrees of freedom. If $D/(n-(p+q+l+3))$ is much larger than one, then the fitted model is found to be inadequate \citep{Myers2010}.

%\subsection{Model selection criteria}

Model selection among several competing models may be based on the usual information criteria. The Akaike information criterion (AIC)~\citep{Akaike1974} is given by
\begin{align}\label{e:maic}
{\rm AIC} = -2 \widehat{\ell} + 2 (p+q+l+3).
\end{align}
Usual information criteria aims at estimating the expected partial log-likelihood and applying a penalty proportional to the number of parameters in the model (in the AIC case, $2(p+q+l+3)$) for the maximized partial log-likelihood function. If the penalty term $2(p+q+l+3)$ in \eqref{e:maic} is replaced by $\log(n)(p+q+l+3)$, we obtain the Schwarz information criterion (BIC) \citep{Schwarz1978}; if it
is replaced by $\log\big(\log(n)\big)(p+q+l+3)$ instead, the \cite{Hannan1979} criterion (HQ) is obtained.

%\subsection{Residuals}

Residual analysis is an important step in verifying whether the estimated model provides a good fit to the data.
Since the proposed model is an extension of the beta regression model~\citep{Ferrari2004},
the residual analysis applied to the former can be also applied to the $\beta$ARFIMA models \citep{espinheira2008, espinheira2008b}.
At the outset, we define the following standardized residual:

\begin{align}\label{eq:sw-res}
\widehat{r}_t^1 = \frac{y_t-\widehat{\mu}_t}{\sqrt{\widehat{\rm Var}(y_t)}} =
\frac{y_t - \widehat{\mu}_t}{\sqrt{\widehat{\mu}_t(1-\widehat{\mu}_t)/(1+\widehat{\nu})}}, \quad \mbox{for } t=1,\dots, n.
\end{align}
A more sophisticated residual is the standardized weighted residual, introduced by \cite{espinheira2008b}, which is given by
\begin{align*}
\widehat{r}_t^w = \frac{y_t^\ast-\widehat{\mu}_t^\ast}{\sqrt{\widehat{\rm Var}(y_t^\ast)}} =
\frac{y_t^\ast-\widehat{\mu}_t^\ast}{\sqrt{ \psi'(\widehat{\mu}_t \widehat{\nu}) + \psi'\big[(1-\widehat{\mu}_t) \widehat{\nu} \big] }}, \quad \mbox{for } t=1,\dots, n.
\end{align*}
The authors have shown that ${\rm Var}(y_t^\ast)= \psi'({\mu}_t {\nu}) + \psi'\big[(1-{\mu}_t) {\nu} \big]$. See also \cite{Li1994}.
%Indeed, in \cite{Rocha2009} it was considered in a model adequacy test.
%Our preview numerical simulations showed that the residual $r_t^w$ is more reliable than the others.
When the fitted model is correct, these residuals are well approximated by the standard normal distribution.

When the model is correctly specified, the residuals should display white noise behavior, i.e., they should follow a zero mean and constant variance uncorrelated process~\citep{Kedem2002}. A visual inspection of the residual plot is an indispensable tool for a first step residual check \citep{Box2008}.
Let $\widehat{r}_{1}^{(\cdot)}, \dots, \widehat{r}_n^{(\cdot)}$ be any type of residual obtained from the fitted model.
The usual estimate for the residual autocorrelation function (ACF) is
\begin{align*}
\widehat{\rho}(h)= \frac{\sum_{t=1}^{n-h}(\widehat{r}_t^{(\cdot)}-\overline{r}^{(\cdot)})(\widehat{r}_{t+h}^{(\cdot)}-\overline{r}^{(\cdot)})}{\sum_{t=1}^{n-h}(\widehat{r}_t^{(\cdot)}-\overline{r}^{(\cdot)})^2}, \quad h=0,1,\dots,
\end{align*}
where $\overline{r}^{(\cdot)} = \frac{1}{n} \sum_{t=1}^{n}\widehat{r}_t^{(\cdot)}$. When $i>1$ and $n$ is sufficiently large, the distribution of $\widehat{\rho}(i)$ is approximately normal with zero mean and variance $1/n$ \citep{Kedem2002, Anderson1942, Box2008}.
Hence, one can apply the usual $\pm 1.96/\sqrt{n}$ as (95\%) confidence bands in ACF plots as a first visual inspection tool for white noise behavior~\citep{Kedem2002}. Since these bounds are usually conservative in finite samples (tighter than they should be), a Ljung-Box test can also be applied. In that case, for large enough $n$, the test statistics will follow its usual distribution, but see the discussion in \cite{Fokianos2004}.

Applying the partial maximum likelihood estimator  in~\eqref{model}, we can obtain $h$ steps ahead predicted values for the observed response $y_t$, which we denote by  $\widehat {y}_{n+h}=\widehat y_n(h)$,
\begin{align*}%\label{e:futuros}
\widehat{y}_{n}(h)= g^{-1}
\bigg(\!
\widehat{\alpha} + \bs{x}_{t-1}'\widehat{\bs{\beta}}
+ \!\sum_{j=1}^p \widehat{\phi}_j\big[g(y_{n+h-j})-\bs{x}_{n+h-j-1}'\widehat{\bs{\beta}}\big]
+ \! \sum_{k=1}^m \widehat{c}_k \big[\widehat{r}_{n+h-k}\big]
\!\bigg),
\end{align*}
where $m$ is the (user chosen) truncation point for the MA$(\infty)$ representation in \eqref{model}, $\widehat{c}_k$ is the quantity in \eqref{cks} evaluated at the PMLE estimates $\widehat{\theta}$ and $\widehat{d}$,
\begin{align*}
\big[g(y_{n+h-j})-\bs{x}_{n+h-j-1}'\widehat{\bs{\beta}}\big]
& =
\begin{cases}
g(\widehat{y}_{n}(h-j))-\bs{x}_{n+h-j-1}'\widehat{\bs{\beta}},&\text{if $j< h$},\\
g(y_{n+h-j})-\bs{x}_{n+h-j-1}'\widehat{\bs{\beta}}, & \text{if $j\geq h$},
\end{cases}
\end{align*}
and
\begin{align*}
\big[\widehat{r}_{t} \big] &=
\begin{cases}
0,&\text{if $t\leq 0$ or $t\geq n+h$},\\
g(y_{t}) - g(\widehat{\mu}_{t}), & \text{if $1\leq t\leq n$},\\
g\big(\widehat{y}_n(t-n)\big) - g(\widehat{\mu}_{t}), & \text{if $n+1\leq t \leq n+h-1$}.
\end{cases}
\end{align*}

Finally, in the absence of covariates or when {each of the covariates forms a} stationary sequence with absolutely summable autocorrelation function, identification of long-range dependence in the conditional mean can be done by using the covariance decay of either $y_t$, $g(y_t)$ or cumulative average as a proxy. A slow ACF decay in any of these sequences indicates the presence of long-range dependence in the conditional mean. In the presence of time varying covariates with either, non-stationary, non absolutely summable ACF  or deterministic behavior, the diagnostic can only be done after dealing with the non-stationarity in the series.

\section{Monte Carlo simulation}

In this section we  present a Monte Carlo simulation study to assess the finite sample properties of the PMLE for $\beta$ARFIMA models
{as well as the LR and Wald's $z$ tests for the presence of long-range dependence. Observe that the $z$ statistics is obtained from the information matrix under the alternative hypothesis, while the LR statistics is directly obtained from the log-likelihood, which favor their use in detriment of the Rao's Score test, which requires matrix inversion and evaluation of the score vector under both hypothesis.}
We simulate 1,000 replicates of a {$\beta$ARFIMA$(1,d,1)$} model restricted to the interval $(0,1)$,
with
{$\phi_1=0.2$, $\theta_1=-0.3$, two values of $\nu \in \{40,120\}$, $d\in\{0.15,0.30,0.45\}$ and sample sizes $n\in\{1000,3000,5000\}$}. We apply the logit as link function and no covariates were included in the simulations.

Given the vector of parameters $\bs\gamma$, to generate a size $n$ sample from the specified $\beta$ARFIMA$(p,d,q)$ process restricted to the interval $(0,1)$, let $m>p$ denote the cutoff point for the infinite sum in \eqref{model}. We start the algorithm by setting $r_t=0$ for all $t\leq m$ and $\mu_t=g^{-1}(\alpha)$, for $t=1,\dots,m$. Second step: for $t=m+1$, we obtain $\eta_t$ through \eqref{model}, then we set $\mu_t=g^{-1}(\eta_t)$ {and update $r_t=g(y_t)-g(\mu_t)$}. Finally, $y_t$ is generated from \eqref{p1}, using any adequate method (such as the inversion method). We iterate the second step for $t=m+1,\dots,n_0+n$, where $n_0>m$ denotes the size of a possible burn in. The desired sample is $y_{n_0+1},\dots,y_{n_0+n}$. If needed, the sample can be rescaled to $(a,b)$ through $\tilde y_t=a+(b-a)y_t$. {We have also performed a pilot simulation study (not shown) to determine the influence of the cutoff point $m$ in parameter estimation. We found that for $m\geq50$, it has negligible impact on the estimated values and that a good compromise between computational speed and accuracy is $m=100$ (used here). For practical purposes, where only a handful of series are analyzed, $m=200$ seems a good choice.}  {All routines were implemented by the authors and are available in {\tt R} language \citep{R2017} upon request. The code for the main tasks of computing the partial score vector and the information matrix were written in FORTRAN 90 by the authors and called from within {\tt R} ($\beta$ARMA models can also be fitted). Optimization is performed by using the so-called L-BFGS-S algorithm \citep{Byrd1995}, which was also implemented in FORTRAN 90 language based on \cite{Zhu1997} and applied without any parameter constraint. We use analytical derivatives in the optimization procedure, given in Section \ref{s:score}. The iterative optimization algorithm requires initialization. The starting values of the constant ($\alpha$),
the covariate parameter ($\bs\beta$)
and the autoregressive ($\bs\phi$) parameters
were obtained from a linear regression with response
 $Y=\big(g(y_{m+1}), g(y_{m+2}), \ldots, g(y_{n})\big)^\prime$ on the design matrix $X$
\begin{align*}
X=
\begin{bmatrix}
1& x_{m1} & x_{m2} & \cdots & x_{mr} & g(y_{m}) & g(y_{m-1}) & \cdots & g(y_{m-p+1}) \\
1& x_{(m+1)1} & x_{(m+1)2} & \cdots & x_{(m+1)r} & g(y_{m+1}) & g(y_{m}) & \cdots & g(y_{m-p+2}) \\
\vdots & \vdots & \vdots & \ddots & \vdots & \vdots & \vdots & \ddots & \vdots \\
1& x_{n1} & x_{n2} & \cdots & x_{nr} & g(y_{n-1}) & g(y_{n-2}) & \cdots & g(y_{n-p}) \\
\end{bmatrix},
\end{align*}
For the parameter $\bs\theta$, the starting values are set to zero and $d$ is started as 0.001. We apply analytic derivatives where the ones obtained through iteration are initialized with zero for non-observed values. }

\begin{table}[t]
\caption{Monte Carlo simulation results for the PMLE estimator based on 1,000 replications. Presented are the mean, percentage relative bias (RB\%),
sample variance (Var)
and mean square error (MSE); for $\nu=40$.} \label{t:simu-nu40}
\centering
\footnotesize
\begin{tabular}{lrrrrr}
\hline
\multicolumn{6}{c}{\bf Scenario 1}\\
\hline
\multirow{2}{*}{Parameters}	&$ \alpha$&	$\phi_1$&	$\theta_1$&	$\nu$ &	$d$ \\
	& 0.050 	& 	0.200	& 	-0.300	& 40	& 0.150\\
\hline
\multicolumn{6}{c}{$n=1000$}\\
\hline
Mean	& $	0.061	$ & $	0.039	$ & $	-0.125	$ & $	40.209	$ & $	0.132	$ \\
%Bias	& $	0.011	$ & $	-0.161	$ & $	0.175	$ & $	0.209	$ & $	-0.018	$ \\
RB	& $	21.816	$ & $	-80.511	$ & $	-58.465	$ & $	0.523	$ & $	-12.121	$ \\
Var	& $	0.001	$ & $	0.151	$ & $	0.168	$ & $	3.559	$ & $	0.005	$ \\
MSE	& $	0.001	$ & $	0.177	$ & $	0.199	$ & $	3.603	$ & $	0.006	$ \\
\hline
\multicolumn{6}{c}{$n=3000$}\\
\hline
Mean	& $	0.057	$ & $	0.106	$ & $	-0.201	$ & $	40.009	$ & $	0.143	$ \\
%Bias	& $	0.007	$ & $	-0.094	$ & $	0.099	$ & $	0.009	$ & $	-0.007	$ \\
RB	& $	13.983	$ & $	-47.082	$ & $	-32.945	$ & $	0.023	$ & $	-4.522	$ \\
Var	& $	0.001	$ & $	0.066	$ & $	0.077	$ & $	1.280	$ & $	0.001	$ \\
MSE	& $	0.001	$ & $	0.074	$ & $	0.086	$ & $	1.280	$ & $	0.001	$ \\
\hline
\multicolumn{6}{c}{$n=5000$}\\
\hline
Mean	& $	0.054	$ & $	0.149	$ & $	-0.247	$ & $	40.029	$ & $	0.146	$ \\
%Bias	& $	0.004	$ & $	-0.051	$ & $	0.053	$ & $	0.029	$ & $	-0.004	$ \\
RB		& $	8.211	$ & $	-25.326	$ & $	-17.647	$ & $	0.073	$ & $	-2.609	$ \\
Var		& $	0.000	$ & $	0.035	$ & $	0.041	$ & $	0.803	$ & $	0.001	$ \\
MSE		& $	0.000	$ & $	0.037	$ & $	0.044	$ & $	0.804	$ & $	0.001	$ \\
\hline
\multicolumn{6}{c}{\bf Scenario 2}\\
\hline
\multirow{2}{*}{Parameters}	&$ \alpha$&	$\phi_1$&	$\theta_1$&	$\nu$ &	$d$ \\
	& 0.050 	& 	0.200	& 	-0.300	& 40	& 0.300\\
\hline
\multicolumn{6}{c}{$n=1000$}\\
\hline
Mean	& $	0.062	$ & $	0.081	$ & $	-0.150	$ & $	40.130	$ & $	0.265	$ \\
%Bias	& $	0.012	$ & $	-0.119	$ & $	0.15	$ & $	0.13	$ & $	-0.035	$ \\
RB		& $	23.743	$ & $	-59.312	$ & $	-50.163	$ & $	0.326	$ & $	-11.709	$ \\
Var		& $	0.005	$ & $	0.198	$ & $	0.201	$ & $	3.679	$ & $	0.006	$ \\
MSE		& $	0.005	$ & $	0.212	$ & $	0.224	$ & $	3.696	$ & $	0.007	$ \\
\hline
\multicolumn{6}{c}{$n=3000$}\\
\hline
Mean	& $	0.059	$ & $	0.156	$ & $	-0.240	$ & $	39.864	$ & $	0.284	$ \\
%Bias	& $	0.009	$ & $	-0.044	$ & $	0.06	$ & $	-0.136	$ & $	-0.016	$ \\
RB		& $	18.641	$ & $	-22.009	$ & $	-20.045	$ & $	-0.339	$ & $	-5.393	$ \\
Var		& $	0.003	$ & $	0.106	$ & $	0.104	$ & $	1.716	$ & $	0.002	$ \\
MSE		& $	0.003	$ & $	0.108	$ & $	0.108	$ & $	1.735	$ & $	0.002	$ \\
\hline
\multicolumn{6}{c}{$n=5000$}\\
\hline
Mean	& $	0.058	$ & $	0.19	$ & $	-0.279	$ & $	39.798	$ & $	0.290	$ \\
%Bias	& $	0.008	$ & $	-0.01	$ & $	0.021	$ & $	-0.202	$ & $	-0.01	$ \\
RB		& $	16.231	$ & $	-4.847	$ & $	-7.106	$ & $	-0.506	$ & $	-3.268	$ \\
Var		& $	0.002	$ & $	0.072	$ & $	0.068	$ & $	1.160	$ & $	0.001	$ \\
MSE		& $	0.002	$ & $	0.072	$ & $	0.068	$ & $	1.201	$ & $	0.002	$ \\
\hline
\multicolumn{6}{c}{\bf Scenario 3}\\
\hline
\multirow{2}{*}{Parameters}	&$ \alpha$&	$\phi_1$&	$\theta_1$&	$\nu$ &	$d$ \\
	& 0.050 	& 	0.200	& 	-0.300	& 40	& 0.450\\
\hline
\multicolumn{6}{c}{$n=1000$}\\
\hline
Mean	& $	0.056	$ & $	0.342	$ & $	-0.349	$ & $	39.838	$ & $	0.359	$ \\
%Bias	& $	0.006	$ & $	0.142	$ & $	-0.049	$ & $	-0.162	$ & $	-0.091	$ \\
RB		& $	12.439	$ & $	70.814	$ & $	16.435	$ & $	-0.406	$ & $	-20.146	$ \\
Var		& $	0.022	$ & $	0.265	$ & $	0.222	$ & $	5.335	$ & $	0.016	$ \\
MSE		& $	0.022	$ & $	0.285	$ & $	0.225	$ & $	5.361	$ & $	0.024	$ \\
\hline
\multicolumn{6}{c}{$n=3000$}\\
\hline
Mean	& $	0.061	$ & $	0.280	$ & $	-0.341	$ & $	39.110	$ & $	0.419	$ \\
%Bias	& $	0.011	$ & $	0.08	$ & $	-0.041	$ & $	-0.89	$ & $	-0.031	$ \\
RB		& $	21.314	$ & $	40.045	$ & $	13.623	$ & $	-2.225	$ & $	-6.935	$ \\
Var		& $	0.042	$ & $	0.162	$ & $	0.142	$ & $	5.492	$ & $	0.004	$ \\
MSE		& $	0.042	$ & $	0.168	$ & $	0.143	$ & $	6.284	$ & $	0.005	$ \\
\hline
\multicolumn{6}{c}{$n=5000$}\\
\hline
Mean	& $	0.060	$ & $	0.264	$ & $	-0.334	$ & $	38.454	$ & $	0.430	$ \\
%Bias	& $	0.01	$ & $	0.064	$ & $	-0.034	$ & $	-1.546	$ & $	-0.02	$ \\
RB		& $	20.494	$ & $	31.897	$ & $	11.325	$ & $	-3.865	$ & $	-4.476	$ \\
Var		& $	0.039	$ & $	0.126	$ & $	0.108	$ & $	9.719	$ & $	0.002	$ \\
MSE		& $	0.039	$ & $	0.130	$ & $	0.110	$ & $	12.109	$ & $	0.003	$ \\

\hline
\end{tabular}
\end{table}
\normalsize

\begin{table}[t]
\caption{Monte Carlo simulation results for the PMLE estimator based on 1,000 replications. Presented are the mean, percentage relative bias (RB\%),
sample variance (Var)
and mean square error (MSE); for $\nu=120$.} \label{t:simu-nu120}
\centering
\footnotesize
\begin{tabular}{lrrrrr}
\hline
\multicolumn{6}{c}{\bf Scenario 1}\\
\hline
\multirow{2}{*}{Parameters}	&$ \alpha$&	$\phi_1$&	$\theta_1$&	$\nu$ &	$d$ \\
	& 0.050 	& 	0.200	& 	-0.300	& 120	& 0.150\\
\hline
\multicolumn{6}{c}{$n=1000$}\\
\hline
Mean	& $	0.062	$ & $	0.022	$ & $	-0.078	$ & $	120.038	$ & $	0.107	$ \\
%Bias	& $	0.012	$ & $	-0.178	$ & $	0.222	$ & $	0.038	$ & $	-0.043	$ \\
RB		& $	23.023	$ & $	-88.796	$ & $	-73.902	$ & $	0.032	$ & $	-28.815	$ \\
Var		& $	0.001	$ & $	0.144	$ & $	0.157	$ & $	29.267	$ & $	0.005	$ \\
MSE		& $	0.001	$ & $	0.176	$ & $	0.206	$ & $	29.269	$ & $	0.007	$ \\
\hline
\multicolumn{6}{c}{$n=3000$}\\
\hline
Mean	& $	0.055	$ & $	0.122	$ & $	-0.218	$ & $	119.735	$ & $	0.144	$ \\
%Bias	& $	0.005	$ & $	-0.078	$ & $	0.082	$ & $	-0.265	$ & $	-0.006	$ \\
RB		& $	10.571	$ & $	-39.087	$ & $	-27.288	$ & $	-0.221	$ & $	-3.819	$ \\
Var		& $	<0.001	$ & $	0.056	$ & $	0.065	$ & $	10.837	$ & $	0.001	$ \\
MSE		& $	<0.001	$ & $	0.063	$ & $	0.072	$ & $	10.907	$ & $	0.001	$ \\
\hline
\multicolumn{6}{c}{$n=5000$}\\
\hline
Mean	& $	0.053	$ & $	0.150	$ & $	-0.248	$ & $	119.646	$ & $	0.147	$ \\
%Bias	& $	0.003	$ & $	-0.050	$ & $	0.052	$ & $	-0.354	$ & $	-0.003	$ \\
RB		& $	6.379	$ & $	-25.146	$ & $	-17.498	$ & $	-0.295	$ & $	-2.044	$ \\
Var		& $	0.000	$ & $	0.033	$ & $	0.039	$ & $	7.142	$ & $	0.001	$ \\
MSE		& $	0.000	$ & $	0.035	$ & $	0.042	$ & $	7.268	$ & $	0.001	$ \\
\hline
\multicolumn{6}{c}{\bf Scenario 2}\\
\hline
\multirow{2}{*}{Parameters}	&$ \alpha$&	$\phi_1$&	$\theta_1$&	$\nu$ &	$d$ \\
	& 0.050 	& 	0.200	& 	-0.300	& 120	& 0.300\\
\hline
\multicolumn{6}{c}{$n=1000$}\\
\hline
Mean	& $	0.059	$ & $	0.076	$ & $	-0.156	$ & $	120.319	$ & $	0.275	$ \\
%Bias	& $	0.009	$ & $	-0.124	$ & $	0.144	$ & $	0.319	$ & $	-0.025	$ \\
RB		& $	17.706	$ & $	-62.087	$ & $	-48.157	$ & $	0.266	$ & $	-8.449	$ \\
Var		& $	0.002	$ & $	0.156	$ & $	0.164	$ & $	33.290	$ & $	0.005	$ \\
MSE		& $	0.002	$ & $	0.172	$ & $	0.185	$ & $	33.392	$ & $	0.006	$ \\
\hline
\multicolumn{6}{c}{$n=3000$}\\
\hline
Mean	& $	0.055	$ & $	0.134	$ & $	-0.228	$ & $	119.989	$ & $	0.292	$ \\
%Bias	& $	0.005	$ & $	-0.066	$ & $	0.072	$ & $	-0.011	$ & $	-0.008	$ \\
RB		& $	10.807	$ & $	-32.772	$ & $	-24.068	$ & $	-0.009	$ & $	-2.781	$ \\
Var		& $	0.001	$ & $	0.066	$ & $	0.071	$ & $	11.275	$ & $	0.001	$ \\
MSE		& $	0.001	$ & $	0.070	$ & $	0.076	$ & $	11.276	$ & $	0.001	$ \\
\hline
\multicolumn{6}{c}{$n=5000$}\\
\hline
Mean	& $	0.055	$ & $	0.154	$ & $	-0.248	$ & $	119.711	$ & $	0.294	$ \\
%Bias	& $	0.005	$ & $	-0.046	$ & $	0.052	$ & $	-0.289	$ & $	-0.006	$ \\
RB		& $	9.142	$ & $	-22.875	$ & $	-17.475	$ & $	-0.241	$ & $	-2.078	$ \\
Var		& $	0.001	$ & $	0.048	$ & $	0.052	$ & $	7.153	$ & $	0.001	$ \\
MSE		& $	0.001	$ & $	0.050	$ & $	0.055	$ & $	7.237	$ & $	0.001	$ \\
\hline
\multicolumn{6}{c}{\bf Scenario 3}\\
\hline
\multirow{2}{*}{Parameters}	&$ \alpha$&	$\phi_1$&	$\theta_1$&	$\nu$ &	$d$ \\
	& 0.050 	& 	0.200	& 	-0.300	&120	& 0.450\\
\hline
\multicolumn{6}{c}{$n=1000$}\\
\hline
Mean	& $	0.045	$ & $	0.361	$ & $	-0.382	$ & $	119.483	$ & $	0.370	$ \\
%Bias	& $	-0.005	$ & $	0.161	$ & $	-0.082	$ & $	-0.517	$ & $	-0.080	$ \\
RB		& $	-9.190	$ & $	80.676	$ & $	27.378	$ & $	-0.431	$ & $	-17.73	$ \\
Var		& $	0.006	$ & $	0.217	$ & $	0.179	$ & $	45.983	$ & $	0.011	$ \\
MSE		& $	0.006	$ & $	0.243	$ & $	0.186	$ & $	46.251	$ & $	0.017	$ \\
\hline
\multicolumn{6}{c}{$n=3000$}\\
\hline
Mean	& $	0.055	$ & $	0.283	$ & $	-0.342	$ & $	117.700	$ & $	0.415	$ \\
%Bias	& $	0.005	$ & $	0.083	$ & $	-0.042	$ & $	-2.3	$ & $	-0.035	$ \\
RB		& $	9.583	$ & $	41.527	$ & $	13.958	$ & $	-1.917	$ & $	-7.839	$ \\
Var		& $	0.007	$ & $	0.144	$ & $	0.124	$ & $	33.723	$ & $	0.003	$ \\
MSE		& $	0.007	$ & $	0.151	$ & $	0.126	$ & $	39.013	$ & $	0.004	$ \\
\hline
\multicolumn{6}{c}{$n=5000$}\\
\hline
Mean	& $	0.062	$ & $	0.230	$ & $	-0.297	$ & $	117.323	$ & $	0.426	$ \\
%Bias	& $	0.012	$ & $	0.03	$ & $	0.003	$ & $	-2.677	$ & $	-0.024	$ \\
RB		& $	24.326	$ & $	14.838	$ & $	-0.848	$ & $	-2.231	$ & $	-5.243	$ \\
Var		& $	0.008	$ & $	0.102	$ & $	0.090	$ & $	29.791	$ & $	0.002	$ \\
MSE		& $	0.008	$ & $	0.103	$ & $	0.090	$ & $	36.959	$ & $	0.002	$ \\
\hline
\end{tabular}
\end{table}
\normalsize

{Tables~\ref{t:simu-nu40} and \ref{t:simu-nu120} present the simulation results for point estimates.} Performance statistics presented are the mean, percentage relative bias (RB\%), variance (Var) and mean square error (MSE).
The percentage relative bias is defined as the ratio between the bias and the true parameter value times 100.
{For $n=1,000$, there is a small bias for parameter $d$, which is expected since, in the context of long-range dependent processes, it is quite common the presence of bias for smaller sample sizes \citep[see, for instance,][and references therein]{Reisen}.
Overall, the results in Table~\ref{t:simu-nu40}, for all $d$'s show somewhat smaller bias for the parameters $\alpha,\nu,d$ and considerably higher bias for the other estimates. As expected, as $n$ increases, the bias in the estimates decrease (except for $\nu$, but the difference is so small that it can be considered negligible) and so are the variance and MSE, which is a reflection of the PLME's consistency.}
%
%{From the results in Table~\ref{t:simu-nu40}, for $d=0.15$, $0.3$ and $0.45$, for the small (for long memory processes) sample size $n=1000$, the results show relatively smaller bias for the estimators of the parameters $\alpha,\nu,d$ and considerably higher bias for the other estimators. For $n=5,000$, only $\phi_1$ and $\theta_1$ still present relative bias over 15\%. The MSE are relatively small for all parameters and cases, decreasing in value as $n$ increases, which provides numeric evidence of its consistency.
%Table~\ref{t:simu-nu120} shows similar results to that in Table~\ref{t:simu-nu40}, with small precision.
%This fact evidences that the performances of the point estimators are almost invariant with respect to the precision parameter value.
%}

{
We also evaluate the performance of the $z$ and LR statistics for testing the null hypothesis $\mathcal{H}_0: \, d=0$
against two-sided alternative hypothesis.
For this purpose we consider three nominal levels: $1\%$, $5\%$ and $10\%$, and the same scenarios described above.
 }

{ Table \ref{T:barfima11-tamanho} presents the null rejection rates of the two different tests. The figures in this table
clearly show that the test based on $z$ statistics is considerably oversized (liberal) in smaller samples.
In the other hand, the LR test presents the best performer,  being much less distorted than $z$ test.
The LR test's null rejection rates are closer to the nominal levels than Wald's.
}

{
We also present the non-null rejection rates, i.e., their estimated power. The results are presented in Table \ref{T:barfima11-poder}.
As expected, the tests become more powerful as $d$ moves away from zero
and as $n$ increases.
We also notice that the $z$ test is more powerful than the LR test. However, the $z$ test is considerably oversized and this can be an unfair comparison.
Therefore, we conclude that the LR test is more reliable to test the presence of long-range dependence in $\beta$ARFIMA model than the $z$ test.
}

\begin{table}%[t]
\caption{Null rejection rates (\%) for the test of $\mathcal{H}_0:\,d=0$.} \label{T:barfima11-tamanho}
%\tablesize
\begin{center}
\begin{tabular}{lll|rrr}
\hline		
$\nu$ & $\alpha$ & \backslashbox[10pt][c]{Stat}{\!\!$n$} & $	1000	$ & $	3000	$ & $	5000	$ \\
\hline
$ 120 $	& $1\%$	& LR	& $	1.4	$ & $	0.7	$ & $	1.2	$ \\
	&	& $z$	& $	6.8	$ & $	4.1	$ & $	3.5	$ \\
	& $5\%$	& LR	& $	5.6	$ & $	4.3	$ & $	6.9	$ \\
	& 	& $z$	& $	14.2	$ & $	7.9	$ & $	9.9	$ \\
	& $10\%$	& LR	& $	10.7	$ & $	11.2	$ & $	11.4	$ \\
	& 	& $z$	& $	19.8	$ & $	13.8	$ & $	15.5	$ \\
	\hline
$ 40 $	& $1\%$	& LR	& $	1.4	$ & $	0.8	$ & $	1.1	$ \\
	&	& $z$	& $	8.2	$ & $	4.9	$ & $	4.3	$ \\
	& $5\%$	& LR	& $	5.1	$ & $	4.9	$ & $	5.2	$ \\
	& 	& $z$	& $	14.9	$ & $	8.4	$ & $	8.0	$ \\
	& $10\%$	& LR	& $	10.9	$ & $	10.0	$ & $	9.8	$ \\
	& 	& $z$	& $	20.5	$ & $	14.8	$ & $	12.4	$ \\
\hline
\end{tabular}
\end{center}
%\end{table}
%
%\begin{table}[t]
\caption{Non-null rejection rates (\%); significance level of $5\%$.} \label{T:barfima11-poder}
%\tablesize
\begin{center}
\begin{tabular}{lll|rrr}
\hline		
$\nu$ & $d$ & \backslashbox[10pt][c]{Stat}{\!\!$n$} & $	1000	$ & $	3000	$ & $	5000	$ \\
\hline
$ 120 $	& $ 0.15	$ &	LR	& $	37.7	$ & $	98.8	$ & $	99.8	$ \\
	&	 &	$z$	& $	60.9	$ & $	99.3	$ & $	100.0	$ \\
	& $ 0.30	$ &	LR	& $	78.8	$ & $	99.2	$ & $	99.8	$ \\
	&	 &	$z$	& $	95.7	$ & $	100.0	$ & $	100.0	$ \\
	& $ 0.45	$ &	LR	& $	79.9	$ & $	97.5	$ & $	99.2	$ \\
	&	 &	$z$	& $	95.2	$ & $	100.0	$ & $	100.0	$ \\
	\hline
$ 40 $	& $ 0.15	$ &	LR	& $	65.7	$ & $	98.2	$ & $	99.6	$ \\
	&	 &	$z$	& $	74.3	$ & $	99.5	$ & $	99.7	$ \\
	& $ 0.30	$ &	LR	& $	70.9	$ & $	97.9	$ & $	99.4	$ \\
	&	 &	$z$	& $	93.8	$ & $	100.0	$ & $	100.0	$ \\
	& $ 0.45	$ &	LR	& $	76.5	$ & $	94.3	$ & $	92.5	$ \\
	&	 &	$z$	& $	96.1	$ & $	99.9	$ & $	99.9	$ \\
\hline
\end{tabular}
\end{center}
\end{table}

\FloatBarrier
\section{Real data application}

The relative air humidity (or simply relative humidity, abbreviated RH) is an important meteorological characteristic to public health, irrigation scheduling design, and hydrological studies. Low RH is known {to cause health problems}, such as allergies, asthma attacks, dehydration, nasal bleeding, among others, while high RH besides causing respiratory problems, {is responsible for} the increase in precipitation which, in excess, can cause serious consequences, such as flooding in urban areas, landslides, damages to agriculture, etc.
\begin{figure}[h!]
\centering
\subfigure[RH time series]{\includegraphics[width=0.7\textwidth]{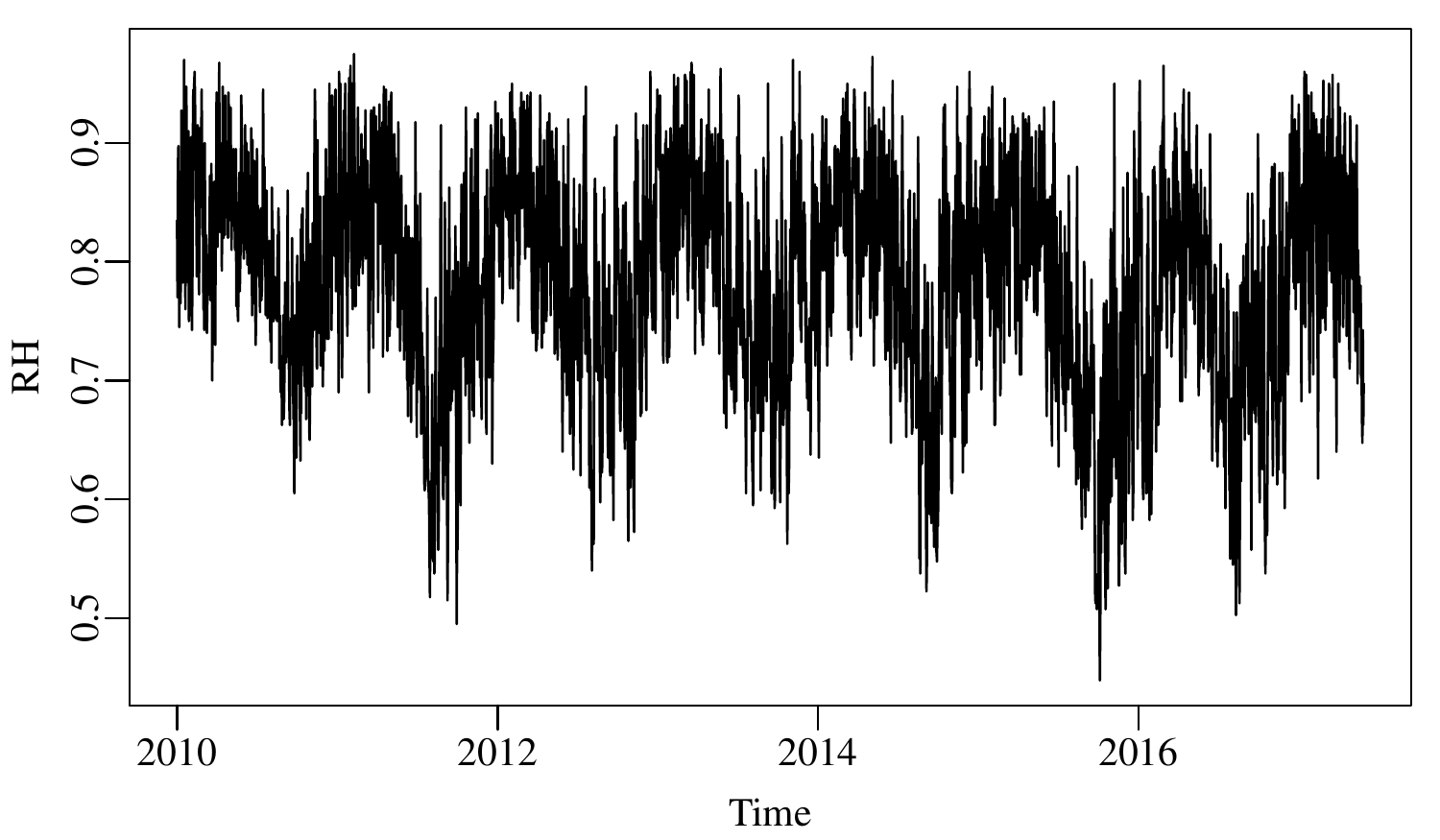}}
\subfigure[ACF]{\includegraphics[width=0.45\textwidth]{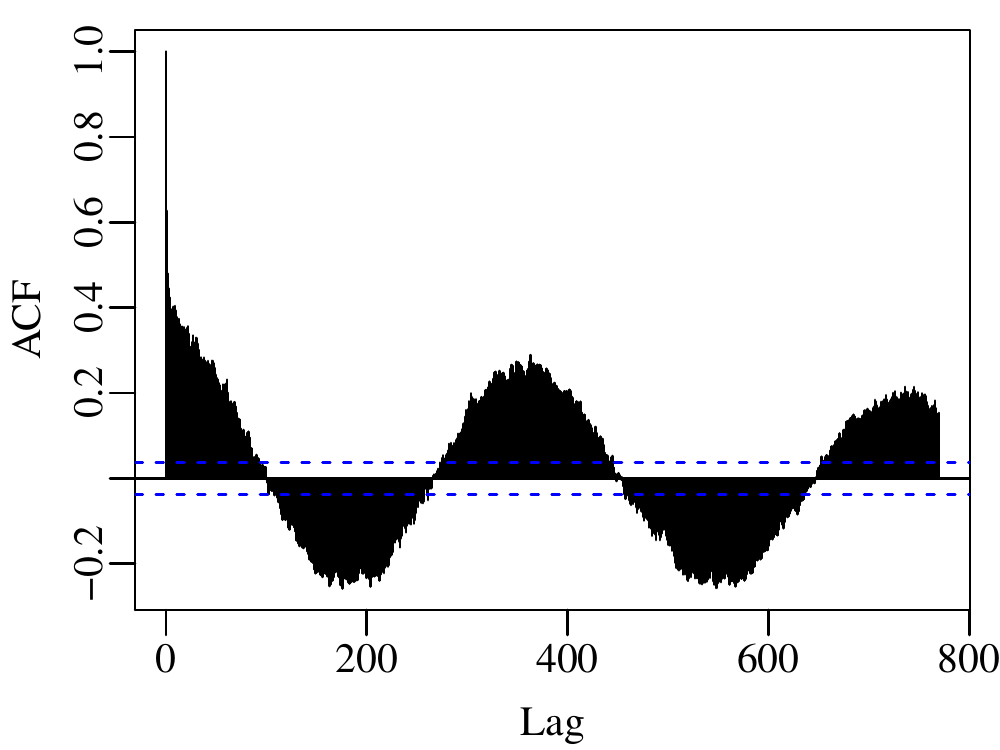}}
\subfigure[Partial ACF]{\includegraphics[width=0.45\textwidth]{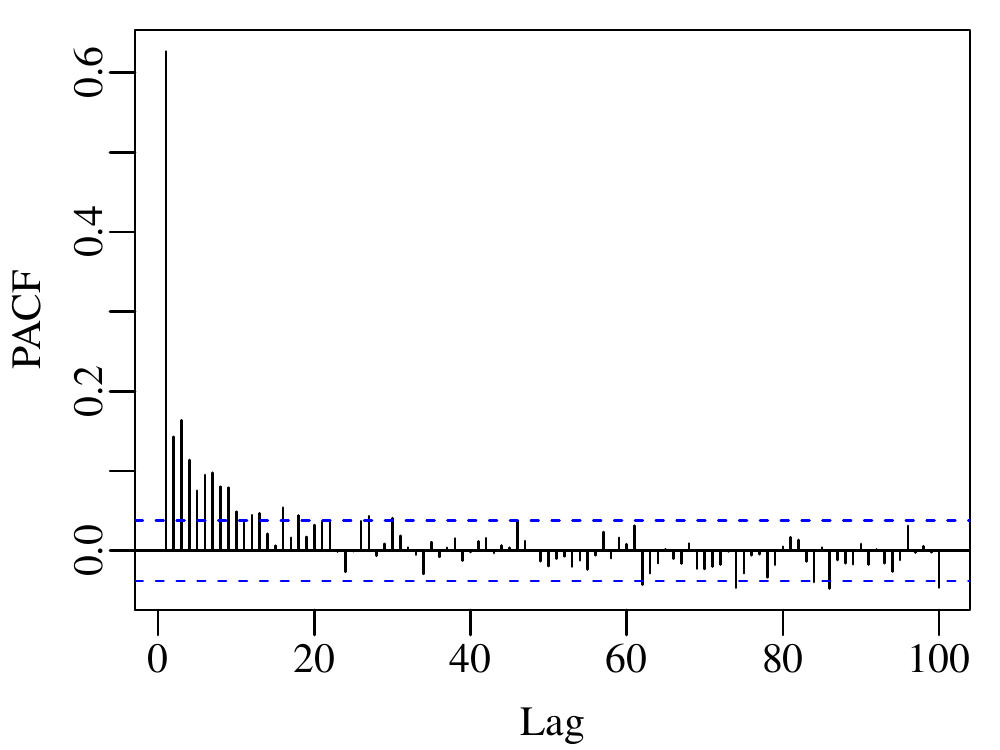}}
\caption{(a) Plor of the daily average RH measured in Manaus, Brazil, from 01/01/2010 to 05/29/2017, the associated (b) sample ACF and (c) partial ACF.}\label{las}
\end{figure}
%\FloatBarrier
To exemplify the usefulness of the proposed $\beta$ARFIMA model, we present an application to the daily average of the RH in Manaus, the Amaz\^onia State capital in Brazil, from 01/01/2010 to 05/29/2017, yielding a sample size of $n=2,704$. The data from 05/30/2016 to 05/29/2017 is reserved to measure the out-of-sample forecast performance of the presented models. The data can be freely obtained from the Instituto Nacional de Meteorologia's (INMET - Brazilian National Institute of Meteorological Research) website (\url{http://www.inmet.gov.br}). The particular station from where the data was collected is situated at longitude $3^\circ$06' south, latitude $60^\circ$ west in Manaus.

Figure \ref{las} presents the RH time series and its autocorrelation function and partial autocorrelation function (PACF). The time series plot reveals a very distinctive seasonality, which we shall incorporate into the  $\beta$ARFIMA model as a covariate. We define this covariate as the seasonal part of a Holt-Winters' decomposition {(additive)} of the time series \citep{Winters1960}. This decomposition is also useful for out-of-sample forecasting as future values for the covariates can be trivially obtained from it. To fit the model, we use $m=200$, the logit as link function  and the diagnostics are based on the standardized residual defined in  \eqref{eq:sw-res}, while $p$-values are obtained from the Wald's $z$ test \eqref{W}. { To select a model to the data, we systematically try different order $\beta$ARFIMA$(p,d,q)$ models and select the one whose parameter are all significant and whose residual does not reject the null hypothesis (using 20 lags) in the Ljung-Box test. All tests are conducted at 5\% significance level. For comparison purposes, we also fit an additive Holt-Winters and a $\beta$ARMA model. To fit and select the $\beta$ARMA model we follow a similar approach as the $\beta$ARFIMA. The routines in {\tt R} and data used in this section are available upon request.}

{Based on the criteria explained above, we have selected a $\beta$ARFIMA$(0,d,1)$ model for the relative humidity data. Table \ref{t:fitted} presents the fitted $\beta$ARFIMA model along with some diagnostics.}
%
%\FloatBarrier
\begin{table}   [h!]
\caption{Fitted $\beta$ARFIMA$(0,d,1)$ model for the relative humidity data in Manaus.}\label{t:fitted}
\begin{center}
\begin{tabular}{crcrc}
\hline
             &Estimate      & Std. Error  & $z$ stat.  & Pr$(>|z|)$ \\
\hline
$\nu$        &  30.2799     &  0.8772     & 34.5207    &  0.0000 \\
$d$          &   0.2869     &  0.0190     & 15.1098    &  0.0000 \\
$\alpha$     &   1.0700     &  0.0526     & 20.3571    &  0.0000 \\
$\beta_1$    &   1.1606     &  0.1904     &  6.0942    &  0.0000 \\
$\theta_1$   &   0.0854     &  0.0268     &  3.1823    &  0.0015 \\
\hline
 \multicolumn{5}{c}{Log-likelihood: $2890.7$}     \\
 \multicolumn{5}{c}{AIC: $-5771.5$ \qquad BIC: $-5742.7$}\\
 \multicolumn{5}{c}{LR test for $\mathcal{H}_0: d=0$ \qquad $p$-value $= 0.000$}\\
 \multicolumn{5}{c}{Ljung-Box test (df = 20)\qquad $p$-value = $0.302$}\\
    \hline
\end{tabular}
\end{center}
\end{table}

{Proceeding similarly as in the $\beta$ARFIMA case, we have selected a $\beta$ARMA(1,2) for the data set. Table \ref{tab:fitted} presents the fitted model along with some diagnostics. Simpler models did present all significative coefficients, but failed the Ljung-Box test. }
\begin{table}   [h!]
\caption{Fitted $\beta$ARMA$(1,2)$ model for the relative humidity data in Manaus.}\label{tab:fitted}
\begin{center}
\begin{tabular}{crrrr}
\hline
             &Estimate  & Std. Error & $z$ stat. & Pr$(>|z|)$\\
             \hline
$\nu$         & 30.5091  &   0.8845  & 34.4950 &  0.0002\\
$\alpha$      & 0.0247   &   0.0067  & 3.6922  &  0.0000\\
$\beta_1$     & 1.0785   &   0.1909  & 5.6491 &  0.0000\\
$\phi_1$      & 0.9757   &   0.0054  & 180.7086 &  0.0000\\
$\theta_1$    & -0.6252  &   0.0201  & 31.0768 &  0.0000\\
$\theta_2$    & -0.2346  &   0.0194  & 12.0943  &  0.0000\\
\hline
 \multicolumn{5}{c}{Log-likelihood: $2903.2$}     \\
 \multicolumn{5}{c}{AIC: $-5794.4$ \qquad BIC: $-5759.9$}\\
 \multicolumn{5}{l}{Ljung-Box test (df = 20)\qquad $p$-value = $0.392$}\\
    \hline
\end{tabular}
\end{center}
\end{table}
{It is well known in the literature that a long-range dependent process can be well approximated by an ARMA process for which the roots of the autoregressive polynomial are close to the unit circle \cite[the case of][is emblematic]{prass}. It is very interesting to notice that this is also reflected in the present case. Observe that the values of $\beta_1$ and $\nu$ on the fitted $\beta$ARMA and $\beta$ARFIMA  are close but the AR part of the fitted $\beta$ARMA model present a root very close to the unitary circle ($\approx1.03$). In terms of model selection,  all goodness of fit criteria (AIC, BIC and log-likelihood) suggest the $\beta$ARMA as the best model for the data, but the difference is almost imperceptible. Also notice that, both Wald's $z$ test and the LR test point to a significant long-range dependence parameter $d$.  }

{Table \ref{tab:hwmodel} present the fitted additive Holt-Winters model. As expected, the trend coefficient is zero as the data presents no trend. For this model a Ljung-Box test shows that there is still serial dependence in the residuals (defined as observed minus fitted values) from the Holt-Winters model ($p$-value $< 2.2\times10^{-16}$). The Holt-Winters is a predictive model so there are no diagnostics for it.
}
\begin{table}[!h]
\begin{center}
  \caption{Fitted additive Holt-Winters model for the relative humidity data in Manaus with additive seasonal cycle.}  \label{tab:hwmodel}%
    \begin{tabular}{ccc}
   \hline
  level & trend & seasonality \\
    \hline
    0.198 & 0.000     & 0.377 \\
    \hline
    \end{tabular}
    \end{center}
\end{table}%

{ We also present an in-sample and out-of-sample  forecasting study based on the fitted models. As mentioned before, we have reserved the data from 05/30/2016 to 05/29/2017 (365 observations) to compare with the out-of-sample forecasts obtained from the models.}

\begin{figure}[h!]
\centering
\subfigure[$\beta$ARFIMA$(0,d,1)$]{	\includegraphics[width=0.57\textwidth]{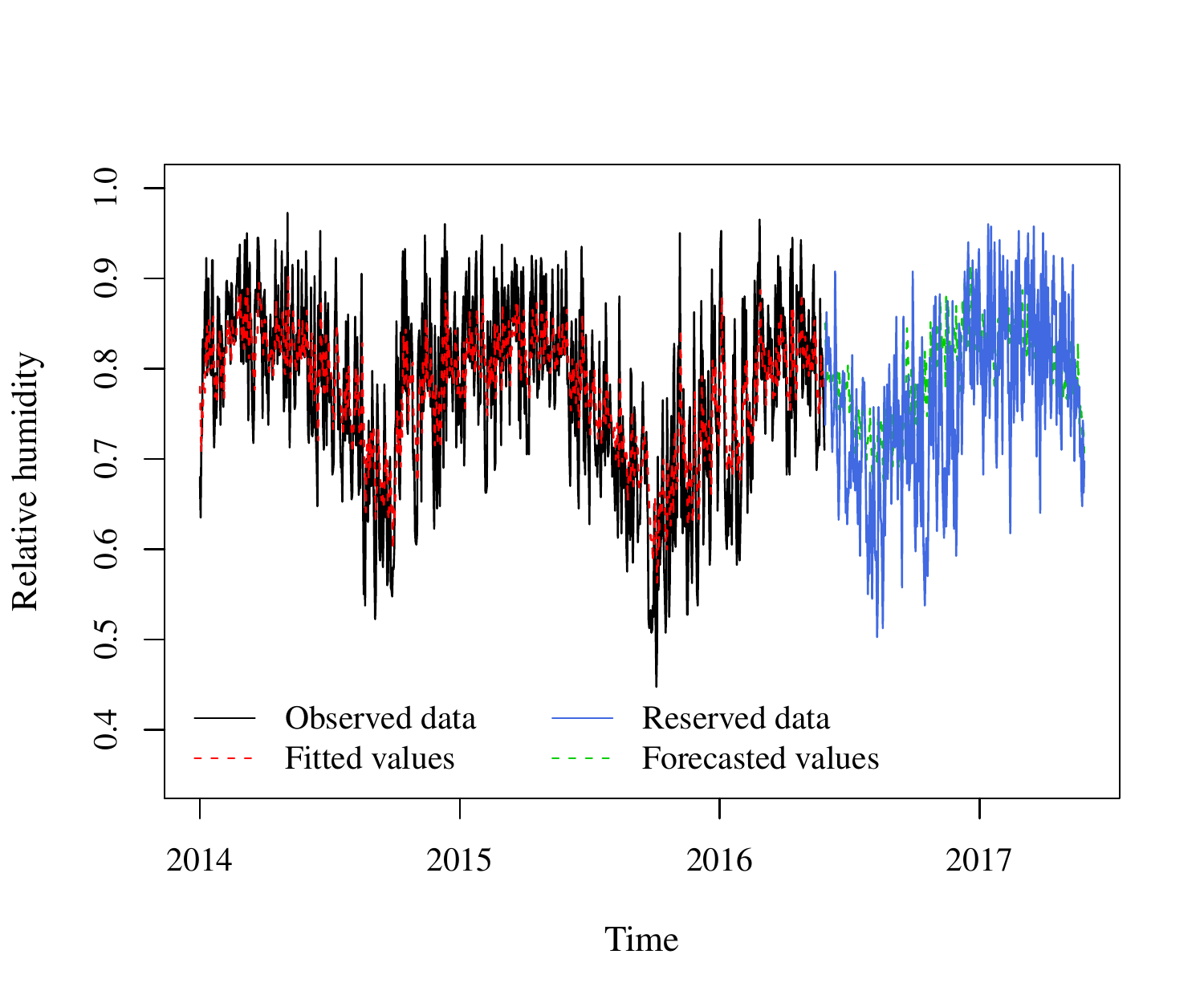}}\vspace{-0.75cm}\\
\subfigure[ $\beta$ARMA$(1,2)$]{\includegraphics[width=0.57\textwidth]{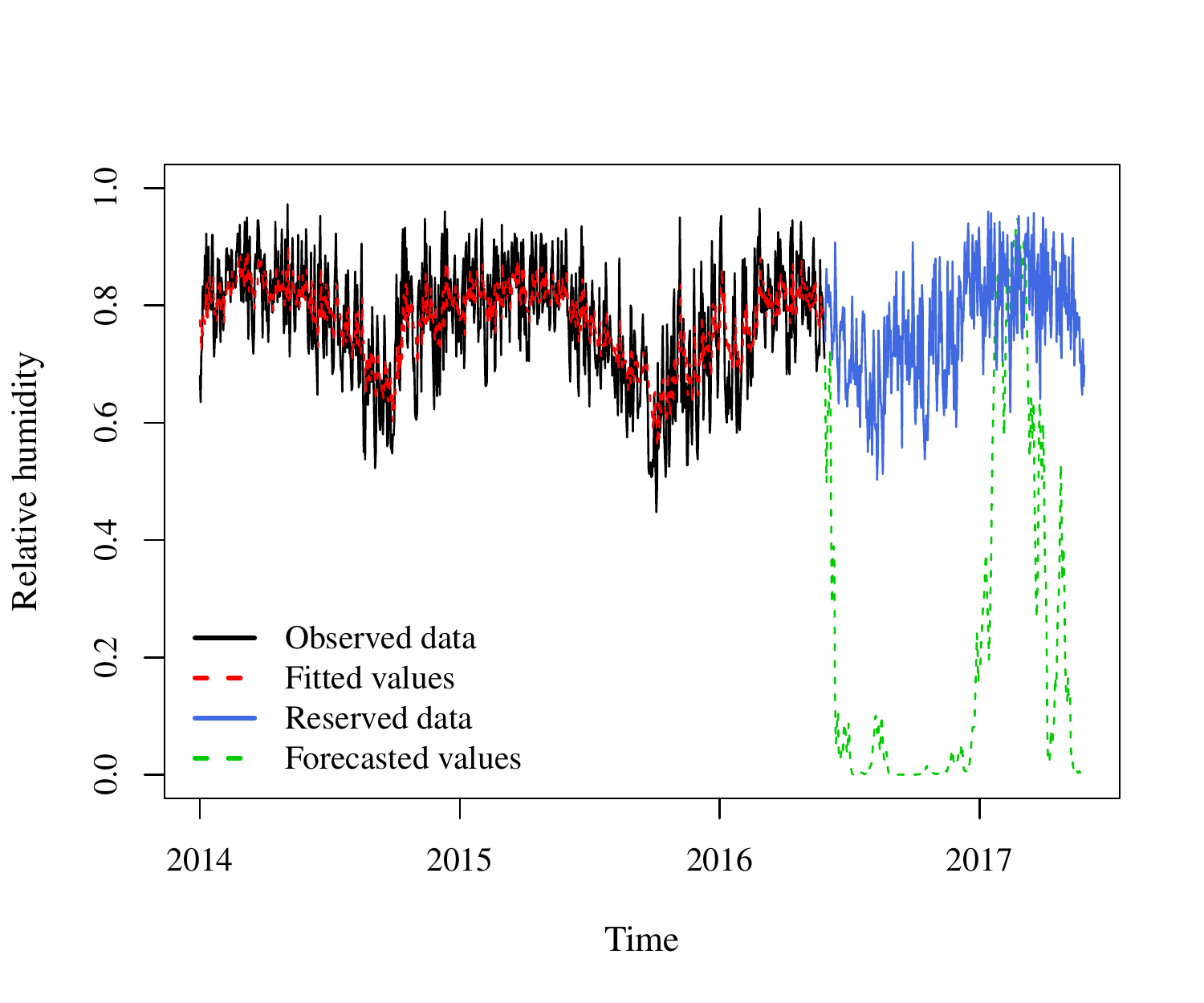}}\vspace{-0.75cm}\\
\subfigure[ Holt-Winters]{\includegraphics[width=0.57\textwidth]{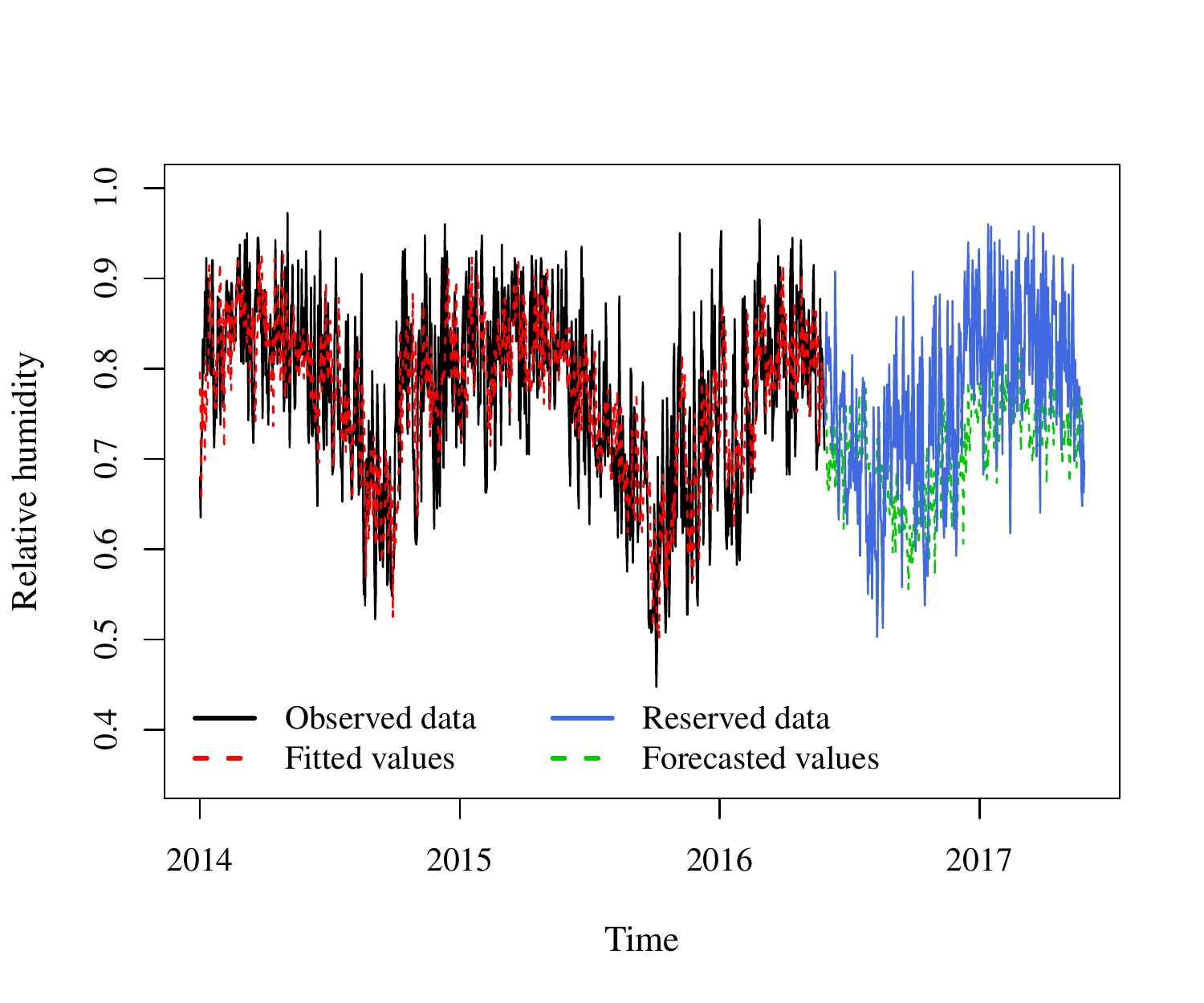}}
\caption{Fitted values, observed and reserved values and 365 step ahead (one year) forecasts for the relative humidity in Manaus.
}\label{f:forecast}
\end{figure}

{The in-sample, 365 steps ahead out-of-sample forecasts for the fitted models, as well as the reserved data  are presented in Figure \ref{f:forecast} while Table \ref{t:forecast} presents some forecasting diagnostics, namely, the root mean squared error (RMSE), the mean absolute error (MAE), and  mean absolute percentage error (MAPE). The out-of-sample diagnostics were obtained from the 365 reserved values, compared to 365 step-ahead forecasts. }

{In terms of in-sample forecast, the plots indicates that both, the $\beta$ARFIMA and $\beta$ARMA models, successfully captured the seasonal component in the data. Also the results presented in Table~\ref{t:forecast} show that the $\beta$ARMA present slightly better in-sample forecast diagnostics compared to the $\beta$ARFIMA model, which is not surprising given that the fitted $\beta$ARMA model presents more parameters. The in-sample forecast for Holt-Winters model seems visually poorer and this is reflected in the diagnostics as well.}

{ Out-of-sample results, however, present a totally different picture. For the $\beta$ARFIMA the out-of-sample forecast seems to predict well the data behavior and the diagnostics are quite good. The out-of-sample forecasts for the $\beta$ARMA are clearly off. This behavior is expected because the AR polynomial in the fitted model present a near-unit root, which induces a near-integrated process behavior in the model's conditional mean.  Hence, even though the $\beta$ARMA presents a slightly better in-sample forecast for the data, slightly better goodness-of-fit measures, the model fails in producing adequate forecasts for the data due to the evidence of long-range dependence in the processes' conditional mean, which is balanced by a near unit root in the AR polynomial. Finally, the Holt-Winters model is capable of producing meaningful out-of-sample forecast, but they are overall poorer when compared to the $\beta$ARFIMA's.}

\begin{table}[h!]
\caption{In-sample and out-of-sample forecasting accuracy measures for the fitted models.}\label{t:forecast}
\begin{center}
\begin{tabular}{cccc}
\hline
   Model                    & RMSE    &     MAE  &   MAPE \\
   \hline
   \multicolumn{4}{c}{In-sample forecasting performance}\\
          \hline
$\beta$ARFIMA$(0,d,1)$ &  0.0690 &  0.0563  &  7.36\% \\
$\beta$ARMA$(1,2)$     &  0.6887 &  0.0559  &  7.29\% \\
Holt-Winters           &  0.0841 &  0.0648  &  8.54\% \\
\hline
\multicolumn{4}{c}{Out-of-sample forecasting performance (365 steps-ahead)}\\
\hline
$\beta$ARFIMA$(0,d,1)$ &  0.0891  & 0.0705  & 9.86\%\\
Holt-Winters           &  0.1703  & 0.1404  & 17.47\%\\
\hline
\end{tabular}
\end{center}
\end{table}

\FloatBarrier

\FloatBarrier
\section{Conclusion and final remarks}

In this work we introduce and study a dynamic time series regression model for bounded continuous random variables observed over time. The proposed model accommodates regressors through a GLM-type structure  and a long-range dependent time series structure. The proposed model generalizes the $\beta$ARMA model of \cite{Rocha2009} by allowing the time series part of the model to present long-range dependence. The model also allows for covariates which can be random, non-random (predetermined) and even time dependent in nature. This flexibility is due to the adopted partial maximum likelihood approach for parameter estimation. Besides introducing the concept in the model's framework, we also explicitly derive the associated score vector and conditional information matrix.

The paper also presents the asymptotic theory for the proposed partial maximum likelihood estimator. We show that the estimator exists and is asymptotically unique, consistent and normally distributed. Based on the asymptotic theory, we provide hypothesis testing, diagnostics and forecast tools for the proposed model.

A Monte Carlo simulation study {is presented and shows the proposed partial maximum likelihood estimator finite sample performance as well as the likelihood ratio and the Wald's $z$ test's. The simulation show an overall good point estimation performance of the PMLE. Regarding the tests, we found that the Wald's $z$ test is considerably oversized, while the LR test performs well in all simulated scenarios. }

{Finally, an application to data regarding daily relative humidity in Manaus, Brazil is presented. We compare the $\beta$ARFIMA, $\beta$ARMA and Holt-Winters models in terms of goodness-of-fit measures as well as in-sample and out-of-sample forecasts. The results show that the $\beta$ARFIMA model was capable of capturing the data dynamics, including seasonality.  Regarding  out-of-sample forecast, the $\beta$ARFIMA was again the best one in terms of commonly used accuracy measures. The fitted $\beta$ARMA  presented a near-unit root in the autoregressive polynomial with severe out-of-sample forecast implications. The application presents a scenario where the $\beta$ARFIMA model is adequate and yields useful forecasts while the $\beta$ARMA, although suitable for the data, fails to produce meaningful forecast due to the evidence of long-range dependence in the data's conditional mean. }

\section*{Acknowledgements}

We gratefully acknowledge partial financial support from CNPq and FAPERGS, Brazil.
We are also grateful to two anonymous referees whose comments and suggestions contributed to improve the paper's quality.

\footnotesize
\bibliographystyle{elsarticle-harv}
\bibliography{betareg}

\begin{thebibliography}{39}
\expandafter\ifx\csname natexlab\endcsname\relax\def\natexlab#1{#1}\fi
\expandafter\ifx\csname url\endcsname\relax
  \def\url#1{\texttt{#1}}\fi
\expandafter\ifx\csname urlprefix\endcsname\relax\def\urlprefix{URL }\fi

\bibitem[{Akaike(1974)}]{Akaike1974}
Akaike, H., 1974. A new look at the statistical model identification. IEEE
  Transactions on Automatic Control 19~(6), 716--723.

\bibitem[{Anderson(1942)}]{Anderson1942}
Anderson, R.~L., 1942. Distribution of the serial correlation coefficient. The
  Annals of Mathematical Statistics 13~(1), 1--13.

\bibitem[{Benjamin et~al.(1998)Benjamin, Rigby, and
  Stasinopoulos}]{Benjamin1998}
Benjamin, M., Rigby, R., Stasinopoulos, D., 1998. Fitting non-{G}aussian time
  series models. COMPSTAT Proceedings in Computational Statistics Heidelburg:
  Physica-Verlag, 191--196.

\bibitem[{Benjamin et~al.(2003)Benjamin, Rigby, and
  Stasinopoulos}]{Benjamin2003}
Benjamin, M., Rigby, R., Stasinopoulos, D., 2003. Generalized autoregressive
  moving average models. Journal of the American Statistical Association
  98~(461), 214--223.

\bibitem[{Box et~al.(2008)Box, Jenkins, and Reinsel}]{Box2008}
Box, G., Jenkins, G.~M., Reinsel, G., June 2008. Time series analysis:
  forecasting and control. Hardcover, John Wiley \& Sons.

\bibitem[{Brockwell and Davis(1991)}]{Brockwell1991}
Brockwell, P.~J., Davis, R.~A., 1991. Time Series: Theory and Methods, 2nd
  Edition. Springer-Verlag.

\bibitem[{Byrd et~al.(1994)Byrd, Lu, Nocedal, and Zhu}]{Byrd1995}
Byrd, R.~H., Lu, P., Nocedal, J., Zhu, C., 1994. A limited-memory algorithm for
  bound constrained optimization. Siam Journal on Scientific Computing 16,
  1190--1208.

\bibitem[{Cox(1975)}]{Cox1975}
Cox, D.~R., 1975. Partial likelihood. Biometrika 62, 69--76.

\bibitem[{Espinheira et~al.(2008{\natexlab{a}})Espinheira, Ferrari, and
  Cribari-Neto}]{espinheira2008b}
Espinheira, P., Ferrari, S. L.~P., Cribari-Neto, F., 2008{\natexlab{a}}. On
  beta regression residuals. Journal of Applied Statistics 35, 407--419.

\bibitem[{Espinheira et~al.(2008{\natexlab{b}})Espinheira, Ferrari, and
  Cribari-Neto}]{espinheira2008}
Espinheira, P.~L., Ferrari, S. L.~P., Cribari-Neto, F., 2008{\natexlab{b}}.
  Influence diagnostics in beta regression. Computational Statistics \& Data
  Analysis 52, 4417--4431.

\bibitem[{Fahrmeir and Kaufmann(1985)}]{Fahrmeir1985}
Fahrmeir, L., Kaufmann, H., 1985. Consistency and asymptotic normality of the
  maximum likelihood estimator in generalized linear models. The Annals of
  Statistics 1~(13), 342--368.

\bibitem[{Ferrari and Cribari-Neto(2004)}]{Ferrari2004}
Ferrari, S. L.~P., Cribari-Neto, F., 2004. Beta regression for modelling rates
  and proportions. Journal of Applied Statistics 31~(7), 799--815.

\bibitem[{Fokianos and Kedem(1998)}]{Fokianos1998}
Fokianos, K., Kedem, B., 1998. Prediction and classification of non-stationary
  categorical time series. Journal of Multivariate Analysis 67, 277--296.

\bibitem[{Fokianos and Kedem(2004)}]{Fokianos2004}
Fokianos, K., Kedem, B., 2004. Partial likelihood inference for time series
  following generalized linear models. Journal of Time Series Analysis 25~(2),
  173--197.

\bibitem[{Granger and Joyeux(1980)}]{Granger1980}
Granger, C., Joyeux, R., 1980. An introduction to long memory time series and
  fractional differencing. Journal of Time Series Analysis 1, 15--30.

\bibitem[{Guolo and Varin(2014)}]{Guolo2014}
Guolo, A., Varin, C., 03 2014. Beta regression for time series analysis of
  bounded data, with application to {C}anada {G}oogle {F}lu {T}rends. The
  Annals of Applied Statistics 8~(1), 74--88.

\bibitem[{Haberman(1977)}]{Haberman1977a}
Haberman, S., 1977. Maximum likelihood estimates in exponential response
  models. Annals of Statistics 5, 815--841.

\bibitem[{Hannan and Quinn(1979)}]{Hannan1979}
Hannan, E.~J., Quinn, B.~G., 1979. The determination of the order of an
  autoregression. Journal of the Royal Statistical Society. Series B 41~(2),
  190--195.

\bibitem[{Honsking(1981)}]{Hosking1981}
Honsking, J., 1981. Fractional differencing. Biometrika 1~(68), 165--176.

\bibitem[{Jacod(1987)}]{Jacod1987}
Jacod, J., 1987. Partial likelihood processes and asymptotic normality.
  Stochastic Processes and its Applications 26, 47--71.

\bibitem[{Jacod(1990)}]{Jacod1990}
Jacod, J., 1990. On partial likelihood. Annales de l'Institut Henri
  Poincar\`{e} Probabilite\`{e} et Statistiques 26, 299--329.

\bibitem[{Kedem and Fokianos(2002)}]{Kedem2002}
Kedem, B., Fokianos, K., 2002. Regression models for time series analysis. John
  Wiley \& Sons.

\bibitem[{Li(1991)}]{Li1991}
Li, W.~K., 1991. Testing model adequacy for some {M}arkov regression models for
  time series. Biometrika 78~(1), 83--89.

\bibitem[{Li(1994)}]{Li1994}
Li, W.~K., 1994. Time series models based on generalized linear models: Some
  further results. Biometrics 50~(2), 506--511.

\bibitem[{Nordberg(1980)}]{Nordberg1980}
Nordberg, L., 1980. Asymptotic normality of maximum likelihood estimators based
  on independent unequally distributed observations in exponential family
  models. Scandinavian Journal of Statistics 7, 27--32.

\bibitem[{Palma(2007)}]{palma2007}
Palma, W., 2007. Long-Memory Time Series: Theory and Methods. Wiley Series in
  Probability and Statistics. Wiley.

\bibitem[{Pawitan(2001)}]{Pawitan2001}
Pawitan, Y., 2001. In All Likelihood: Statistical Modelling and Inference Using
  Likelihood. Oxford Science publications.

\bibitem[{Prass et~al.(2012)Prass, Bravo, Clarke, Collischonn, and
  Lopes}]{prass}
Prass, T.~S., Bravo, J.~M., Clarke, R.~T., Collischonn, W., Lopes, S. R.~C.,
  2012. Comparison of forecasts of mean monthly water level in the paraguay
  river, brazil, from two fractionally differenced models. Water Resources
  Research 48~(5), w05502.

\bibitem[{{R Core Team}(2017)}]{R2017}
{R Core Team}, 2017. R: A Language and Environment for Statistical Computing. R
  Foundation for Statistical Computing, Vienna, Austria.
\newline\urlprefix\url{https://www.R-project.org/}

\bibitem[{Reisen et~al.(2001)Reisen, Abraham, and Lopes}]{Reisen}
Reisen, V., Abraham, B., Lopes, S. R.~C., 2001. Estimation of parameters in
  {ARFIMA} processes: a simulation study. Communications in Statistics -
  Simulation and Computation 30~(4), 787--803.

\bibitem[{Rocha and Cribari-Neto(2009)}]{Rocha2009}
Rocha, A.~V., Cribari-Neto, F., 2009. Beta autoregressive moving average
  models. Test 18~(3), 529--545.

\bibitem[{Rocha and Cribari-Neto(2017)}]{Rocha2017}
Rocha, A.~V., Cribari-Neto, F., 2017. Erratum to: Beta autoregressive moving
  average models. TEST 26~(2), 451--459.

\bibitem[{Schwarz(1978)}]{Schwarz1978}
Schwarz, G., 1978. Estimating the dimension of a model. The Annals of
  Statistics 6~(2), 461--464.

\bibitem[{Wang(2012)}]{Wang2012}
Wang, X.-F., 2012. Joint generalized models for multidimensional outcomes: A
  case study of neuroscience data from multimodalities. Biometrical Journal 54,
  264--280.

\bibitem[{Winters(1960)}]{Winters1960}
Winters, P., 1960. Forecasting sales by exponentially weighted moving averages.
  Management Science 6, 324--342.

\bibitem[{Wong(1986)}]{Wong1986}
Wong, W., 1986. Theory of partial likelihood. The Annals of Statistics 14,
  88--123.

\bibitem[{Zeger and Qaqish(1988)}]{Zeger1988}
Zeger, S.~L., Qaqish, B., 1988. Markov regression models for time series: A
  quasi-likelihood approach. Biometrics 44~(4), 1019--1031.

\bibitem[{Zhu et~al.(1997)Zhu, Byrd, Lu, and Nocedal}]{Zhu1997}
Zhu, C., Byrd, R.~H., Lu, P., Nocedal, J., 1997. Algorithm 778: L-bfgs-b:
  Fortran subroutines for large-scale bound-constrained optimization. ACM
  Trans. Math. Softw. 23~(4), 550--560.

\bibitem[{Zou et~al.(2010)Zou, Carlsson, and Quinn}]{Zou2010}
Zou, K., Carlsson, M., Quinn, S., 2010. Beta-mapping and beta-regression for
  changes of ordinal-rating measurements on likert scales: A comparison of the
  change scores among multiple treatment groups. Statistics and Medicine 29,
  2486--2500.

\end{thebibliography}

\end{document}